\documentclass[reprint,notitlepage,superscriptaddress,preprintnumbers,nofootinbib,amsmath,amssymb,aps,prd,nobibnotes]{revtex4-1}
\usepackage{graphicx}
\usepackage{dcolumn}
\usepackage{bm}
\usepackage{hyperref}

\newcommand{\uoe}{School of Physics and Astronomy,
University of Edinburgh, Edinburgh EH9 3JZ, United Kingdom}
\newcommand{\bnl}{Physics Department, Brookhaven National Laboratory, Upton, New York 11973, USA}
\newcommand{\cern}{CERN, Theoretical Physics Department, Geneva, Switzerland}
\newcommand{\cambridge}{DAMTP, University of Cambridge, Centre for Mathematical Sciences, Wilberforce Road, Cambridge, CB3 0WA, United Kingdown}
\newcommand{\rccs}{RIKEN Center for Computational Science, Kobe 650-0047, Japan}

\newcommand{\AllRho}{CP-PACS:2007wro,CS:2011vqf,Feng:2010es,Lang:2011mn,Pelissier:2012pi,Dudek:2012xn,Wilson:2015dqa,Bali:2015gji,Bulava:2016mks,Fu:2016itp,Andersen:2018mau,Erben:2019nmx,Alexandrou:2017mpi,ExtendedTwistedMass:2019omo,Fischer:2020yvw}
\newcommand{\AllKstar}{Fu:2012tj,Prelovsek:2013ela,Wilson:2014cna,Wilson:2019wfr,Bali:2015gji,Brett:2018jqw,Rendon:2020rtw}
\newcommand{\AllFV}{Luscher:1985dn,Luscher:1986pf,Luscher:1990ux,Rummukainen:1995vs,Kim:2005gf,Fu:2011xz,Leskovec:2012gb,Bernard:2010fp,Doring:2011vk,Briceno:2012yi,Hansen:2012tf,Briceno:2014oea}
\newcommand{\QCDResReviews}{Briceno2018,Mai:2022eur}

\newcommand{\TwoToThree}{Briceno:2017tce}
\newcommand{\ThreeParticles}{Hansen:2019nir,Mai:2021lwb,Romero-Lopez:2022usb}

\DeclareMathOperator{\valueainverse}{ 1.7295(38)~\mathrm{GeV} } 
\DeclareMathOperator{\valueourpionmass}{ 138.5(2)~\mathrm{MeV} } 
\DeclareMathOperator{\valueourkaonmass}{ 498.9(4)~\mathrm{MeV} }

\usepackage[dvipsnames]{xcolor}
\usepackage{color}
\usepackage{amsmath,amssymb,braket,slashed,mathrsfs,dsfont}
\usepackage{pifont}
\usepackage{braket}
\usepackage[compat=1.1.0]{tikz-feynhand}
\usepackage{mathtools}
\usepackage{subcaption}
\usepackage[mathscr]{euscript}
\usepackage{multirow}
\usepackage[normalem]{ulem} 

\hypersetup{colorlinks, linkcolor = [rgb]{0,0.0,0.75}, citecolor = [rgb]{0,0.0,0.75}, urlcolor = [rgb]{0,0.0,0.75}}

\newcommand{\cf}{\textit{cf.}~}
\newcommand{\eg}{\textit{e.g.}~}
\newcommand{\ie}{\textit{i.e.}~}

\DeclareMathOperator{\Nvec}{\mathnormal{N}_{\mathrm{vec}}}

\DeclareMathOperator{\SNRmin}{{\mathrm{SNR}}_{\mathrm{min}}}
\DeclareMathOperator{\deltatmin}{\mathnormal{\delta t}_{\mathrm{min}}}
\DeclareMathOperator{\Nscan}{\mathnormal{N}_{\mathrm{scan}}}
\DeclareMathOperator{\dmf}{\mathnormal{\mathcal{M}}}
\DeclareMathOperator{\tr}{\mathrm{tr} }
\DeclareMathOperator{\mev}{\!~MeV} 
\DeclareMathOperator{\gev}{\!~GeV} 

\DeclareMathOperator{\AIC}{\mathrm{AIC}}

\newcommand{\runs}{{\mathrm{runs}}}
\newcommand{\run}[1]{\mathrm{run}\text{#1}}

\newcommand{\PS}{{\mathrm{PS}}}
\newcommand{\corr}{{\mathrm{corr}}}
\newcommand{\cm}{{\mathrm{cm}}}
\newcommand{\mdl}{{\mathrm{mod}}}
\newcommand{\BW}{{\mathrm{BW}}}
\newcommand{\ERE}{{\mathrm{ERE}}}
\newcommand{\stat}{{\mathrm{stat}}}
\newcommand{\sys}{{\mathrm{sys}}}
\newcommand{\dd}{{\mathrm{dd}}}
\newcommand{\oth}{{\mathrm{other}}}
\newcommand{\scale}{{\mathrm{scale}}}

\newcommand{\tstart}{t_{\mathrm{start}}}
\newcommand{\tstop}{t_{\mathrm{stop}}}
\newcommand{\tmin}{t_\mathrm{min}}
\newcommand{\tmax}{t_\mathrm{max}}
\newcommand{\AICc}{\mathrm{AIC}_\mathrm{corr}}
\newcommand{\AICps}{\mathrm{AIC}_\mathrm{PS}}
\newcommand{\AICt}{\mathrm{AIC}_\mathrm{t}}

\newcommand{\CovEig}{\bm{\Sigma}_{{\mathrm{eig}},i}}
\newcommand{\CovPS}{\bm{\Sigma}_{\mathrm{PS}}}

\newcommand{\tmat}[1]{t_{#1}} 

\makeatletter
\newcommand{\thickhline}{%
    \noalign {\ifnum 0=`}\fi \hrule height 1pt
    \futurelet \reserved@a \@xhline
}
\newcolumntype{"}{@{\hskip\tabcolsep\vrule width 1pt\hskip\tabcolsep}}
\makeatother

\begin{document}

\preprint{CERN-TH-2024-087}
\title{Physical-mass calculation of $\rho(770)$ and $K^*(892)$ resonance parameters\\ via $\pi \pi$ and $K \pi$ scattering amplitudes from lattice QCD}

\author{Peter Boyle}\affiliation{\bnl}\affiliation{\uoe}
\author{Felix Erben}\affiliation{\cern}\affiliation{\uoe}
\author{Vera G\"ulpers}\affiliation{\uoe}
\author{Maxwell T. Hansen}\affiliation{\uoe}
\author{Fabian Joswig}\affiliation{\uoe}
\author{\\Michael Marshall}\affiliation{\uoe}
\author{Nelson Pitanga Lachini}\email[e-mail: ]{np612@cam.ac.uk}\affiliation{\cambridge}\affiliation{\uoe}
\author{Antonin Portelli}\affiliation{\uoe}\affiliation{\cern}\affiliation{\rccs}

\date{\today}

\begin{abstract}
    We present our study of the $\rho(770)$ and $K^*(892)$ resonances from lattice quantum chromodynamics~(QCD) employing domain-wall fermions at physical quark masses. We determine the finite-volume energy spectrum in various momentum frames and obtain phase-shift parametrizations via the Lüscher formalism, and as a final step the complex resonance poles of the $\pi \pi$ and $K \pi$ elastic scattering amplitudes via an analytical continuation of the models. 
    By sampling a large number of representative sets of underlying energy-level fits, we also assign a systematic uncertainty to our final results. This is a significant extension to data-driven analysis methods that have been used in lattice QCD to date, due to the two-step nature of the formalism. 
    Our final pole positions, $M+i\Gamma/2$, with all statistical and systematic errors exposed, are 
    $M_{K^{*}} = 893(2)(8)(54)(2)\mev$
    and 
    $\Gamma_{K^{*}} = 51(2)(11)(3)(0)\mev$
    for the $K^*(892)$ resonance and 
    $M_{\rho} = 796(5)(15)(48)(2)\mev$
    and 
    $\Gamma_{\rho} = 192(10)(28)(12)(0)\mev$
    for the $\rho(770)$ resonance. The four differently grouped sources of uncertainties are, in the order of occurrence: statistical, data-driven systematic, an estimation of systematic effects beyond our computation (dominated by the fact that we employ a single lattice spacing), and the error from the scale-setting uncertainty on our ensemble.
\end{abstract}

\maketitle

\section{Introduction}
\label{sec:introduction}

Over the last two decades, lattice quantum chromodynamics (QCD) has made significant contributions to the field of hadron spectroscopy, including major developments in calculations involving QCD resonances.\footnote{For recent reviews see, for example, Refs.~\citep{\QCDResReviews}.} While properties of states that are stable under the strong interaction, \eg~pions and kaons, can often be accessed directly from Euclidean lattice correlation functions, the treatment of resonances requires a more intricate formalism. A particularly successful strategy is to relate the finite-volume energy spectrum to hadronic scattering amplitudes~\citep{\AllFV}, which can then be used to extract QCD resonance parameters by analytically continuing in the energy in order to identify poles in the complex plane.

This work focuses on the application of this method to two QCD resonances, the $\rho(770)$ and the $K^*(892)$. Both are of direct physical interest, and the field of lattice QCD has progressed substantially in providing reliable QCD predictions, \eg~of masses and decay widths, which can be compared to the same parameters extracted from experimental data. In addition, precise knowledge of these resonances and their corresponding hadronic scattering amplitudes is a necessary step towards studying weak and electromagnetic decay and transition amplitudes in which both the $\rho(770)$ and the $K^*(892)$ can appear in both intermediate and final states. Examples that have already been studied using lattice QCD include the ${\pi\gamma \to \pi\pi}$~\citep{Briceno:2016kkp,Alexandrou:2018jbt,Niehus:2021iin} and ${K\gamma \to K\pi}$~\citep{Radhakrishnan:2022ubg} transition amplitudes.

More challenging observables include amplitudes for heavy meson decays like ${B \to \rho (\to \pi\pi) \ell \bar{\nu}}$\footnote{See Refs.~\citep{Leskovec:2022ubd, Leskovec:2024pzb} for recent progress in a lattice calculation of this decay amplitude.} and ${B \to K^* (\to K\pi) \ell^+ \ell^-}$. In particular, ${b \to s \ell^+ \ell^-}$ processes have attracted a great deal of recent interest due to tensions between experimental data and Standard Model predictions, sometimes referred to as $B$ anomalies. Such tensions can be studied directly at the level of branching fractions, \eg~for ${B \to K^{*} \ell^+ \ell^-}$, or in derived quantities like $R_{K^*}$, defined as a ratio of branching fractions with electron and muon final states, as detailed in Ref.~\citep{LHCb:2021trn}. While the tension between Standard Model predictions and experimental data in $R_{K^*}$ has recently reduced with an experimental update~\citep{LHCb:2022vje}, the situation for the individual branching fractions is less clear and reliable predictions are urgently needed~\citep{LHCb:2016ykl,Gubernari:2023puw}.

The $\rho(770)$ resonance appears in the $I=1$ channel of $\pi\pi$ scattering and has angular momentum ($J$) and parity ($P$) given by $J^P = 1^-$. It almost exclusively decays into two pions (spin-zero, pseudoscalars) in a $P$-wave (orbital angular momentum $\ell = 1$), with the branching fraction to this channel deviating from 100\% at the sub-percent level~\citep{PDG2024}. The next most relevant decay channels that have been detected experimentally include the electromagnetic channel mentioned above for the charged states (${\rho^\pm \to \pi^\pm \gamma}$) as well as the three-particle decay for the neutral state ${\rho^0 \to \pi^+ \pi^- \gamma}$. Further final states, including $\pi\eta$ and $\pi\pi\pi\pi$ have yet to be measured but have experimentally set upper bounds well below ${1\%}$. The $K^*(892)$ resonance appears in the $I=1/2$ channel of $K\pi$ scattering, also with $J^P = 1^-$. Its branching ratio to $K\pi$ in a $P$-wave also deviates from 100\% only at the percent level~\citep{PDG2024}, with the next relevant channels being $K\gamma$ and $K\pi\pi$. Because this work is exclusively concerned with the isospin channels mentioned in this paragraph, we will drop the isospin index from all subsequent notation. In particular, when we talk about $\pi\pi$ scattering, we always mean $\pi\pi$ scattering in the $I=1$ channel, and equivalently $K \pi$ denotes the $K\pi$ in the $I=1/2$, unless explicitly mentioned otherwise. 

The single-channel dominance in $\rho(770)$ and $K^*(892)$ decays, and more generally the near-elasticity of ${\pi \pi \to \pi \pi}$ and ${K \pi \to K \pi}$ scattering for center-of-mass (c.m.) energies up to ${\sim 1\gev}$, make these particularly clean single-channel resonances to study using finite-volume energies determined via lattice QCD. To apply the approach, multiple correlation functions are needed, constructed using a large operator set. In this work, we achieve this in a cost-effective manner using the distillation technique~\citep{Peardon:2009gh,Morningstar:2011ka}, which conveniently factorizes the lattice quark propagators, enabling the efficient construction of many correlation functions. The distillation method has been successfully applied in various previous studies of hadronic resonances, typically with higher-than-physical quark masses. Many such studies considered the channels of this work: ${\pi\pi \to \rho \to \pi \pi}$~\citep{\AllRho} and ${K\pi \to K^* \to K \pi}$~\citep{\AllKstar}.

In recent years, many lattice QCD calculations have been performed at physical light-quark masses. The effect of increasing the masses has been studied using models and variants of chiral perturbation theory~\citep{Hanhart:2008mx, Nebreda2010, Bolton:2015psa, Niehus:2020gmf,Yu:2023xxf} and also directly using lattice QCD \citep{Wilson:2019wfr}. While extrapolation is possible, the non-trivial dependence of resonance parameters with varying quark masses can only be described with some level of modeling or a truncation in an effective-field-theory expansion. This necessarily breaks down when sufficiently high precision is reached. Therefore, also with an eye towards precision weak decay determinations, a lattice result directly at the physical light-quark mass is both timely and desirable.

This work presents the first lattice QCD determination of the scattering amplitudes and resonance pole positions for the $\rho(770)$ and the $K^*(892)$ with physical values for light- and strange-quark masses, and with both light and strange quarks dynamically included ($N_f = 2+1$). This is also the first calculation of its kind to perform a data-driven estimation of the systematic uncertainty by varying the fit range used to extract the energies, followed by a model average of different scattering amplitude parametrizations to reach the final observables.

The rest of this paper is organized as follows. In Sec.~\ref{sec:strategy}, we explain our strategy to extract finite-volume energy levels from lattice QCD, which includes the construction of suitable interpolators for states in the finite-volume spectrum. In Sec.~\ref{sec:setup}, we present the computational details of this work by introducing the lattice ensemble, as well as giving details of our distillation setup, the resulting pion and kaon dispersion relations, and the tuning of the respective smearing parameters. We present our analysis methods in Sec.~\ref{sec:analysis}, with an emphasis on the propagation of systematic errors in a two-step fitting procedure. Our results are presented and discussed in Sec.~\ref{sec:results}, and finally, we conclude and give an outlook into possible future applications and the impact of this work in Sec.~\ref{sec:conclusions}. In addition to this long-format manuscript, we have prepared a letter summarising the essential methods and results~\citep{our-letter}. All correlator data produced for this work was made available \citep{boyle_2024_vy9x7-bzn92}.

\section{Lattice Strategy}
\label{sec:strategy}

This work uses a finite-volume scattering formalism, first developed by Lüscher~\citep{Luscher:1985dn,Luscher:1986pf,Luscher:1990ux} for two identical particles at rest, and subsequently extended by various authors to nonzero momentum in the finite-volume frame~\citep{Rummukainen:1995vs,Kim:2005gf}, different particle masses and momenta~\citep{Fu:2011xz,Leskovec:2012gb}, multi-channel systems~\citep{Bernard:2010fp,Doring:2011vk,Briceno:2012yi,Hansen:2012tf} and particles with arbitrary intrinsic spin~\citep{Briceno:2014oea}. The basic strategy of our lattice calculation, following this finite-volume method, can be summarized as follows. In Sec.~\ref{sec:operatorprojection}, we define a set of bilinear and two-bilinear interpolators in the relevant channel, and in Sec.~\ref{sec:distillation} we compute all possible correlation functions within this set of interpolators. In Sec.~\ref{sec:gevp} a matrix of these correlation functions can then be diagonalised, giving us access to the finite-volume energy spectrum. In Sec.~\ref{sec:phaseshift}, we overview the Lüscher quantization condition used in this work.

\subsection{Lattice interpolators}
\label{sec:operatorprojection}

To reliably determine the finite-volume spectra in the $\rho$ and $K^*$ channels, we must first identify a set of interpolators with significant overlap to all states in the energy range of interest. The set used in this work includes the vector bilinears
\begin{align}
\begin{split}
    O_{\rho^+}(x) & = - \bar d(x) \bm{\gamma} \, u(x) \,,  \\
    O_{K^{*+}}(x) & = \bar s(x) \bm{\gamma} \, u(x) \,,
    \label{rhokstarinterpolators}
\end{split}
\end{align}
where $u(x),d(x),s(x)$ are the quark fields of an up, down, and strange quark, respectively, on the lattice site $x$, the barred versions are the corresponding anti-quark fields and $\bm \gamma = (  \gamma_x, \gamma_y, \gamma_z )$ is a vector of the spatial Dirac matrices. The subscripts $\rho^+$ and $K^{*+}$ allude to the quark-model content of the corresponding interpolators. While it is important to include these bilinear interpolators to extract the correct spectrum, it has been empirically observed that not even a single finite-volume energy can be reliably extracted if these operators are used in isolation~\citep{Dudek2010,Thomas2012}.

To circumvent this issue, we also include the two-bilinear interpolators of the form~\footnote{We omit normalization factors in $O_{\pi\pi}, O_{K\pi}$ and $O_{\pi^0}$ for simplicity here, but explicitly use them to get the diagrams in Appendix~\ref{apx:wick}.}
\begin{align}
\begin{split}
    O_{\pi\pi}(x,y) &= O_{\pi^+}(x) O_{\pi^0}(y) -   O_{\pi^0}(x) O_{\pi^+}(y),  \\
    O_{K\pi}(x,y) &= - O_{K^+}(x) O_{\pi^0}(y) + \sqrt{2} \, O_{K^0}(x) O_{\pi^+}(y),
    \label{pipikpiinterpolators}
\end{split}
\end{align}
where the relative factors in the two terms on the right-hand sides of the equations effect the projection to isospins $I=1$ and $I=1/2$, respectively, and where the lattice 4-vectors $x=(\mathbf x,t)$ and $y=(\mathbf y,t)$ have the same temporal component. The pseudoscalar bilinears used above are defined as
\begin{align}
\begin{split}
    O_{\pi^+}(x) &= -\bar d(x) \gamma_5 u(x), \\
    O_{\pi^0}(x) &=  \bar u(x) \gamma_5 u(x) - \bar d(x) \gamma_5 d(x) , \\
    O_{K^+}(x) &= \bar s(x) \gamma_5 u(x), \\
    O_{K^0}(x) &= \bar s(x) \gamma_5 d(x),
    \label{pionkaoninterpolators}
\end{split}
\end{align}
where we have introduced the Dirac matrix $\gamma_5$, required to ensure that the operators are parity negative.

All operators are projected to definite spatial momentum within the two-point correlation functions used to extract the spectrum. When used in isolation, the bilinears in Eqs.~\eqref{rhokstarinterpolators} and \eqref{pionkaoninterpolators} are projected to total momentum~$\mathbf P$ as follows
\begin{align}
    \widetilde{O}_{V} (\mathbf P, t) = a^3 \sum_{\mathbf{x}} e^{-i \mathbf{P} \cdot \mathbf{x}} O_{V} (x)  \,, \\
    \widetilde{O}_{M} (\mathbf P, t) = a^3 \sum_{\mathbf{x}} e^{-i \mathbf{P} \cdot \mathbf{x}} O_{M} (x) \,,
    \label{singlebilinearmomproj}
\end{align}
where $V \in \{ \rho^+ , K^{*+} \}$ and $M \in \{ \pi^+, \pi^0, K^+, K^0 \}$. When used in the context of Eq.~\eqref{pipikpiinterpolators}, the individual bilinears are separately projected with momenta $\mathbf p_1, \mathbf p_2$, satisfying $\mathbf p_1 + \mathbf p_2 = \mathbf P$:
\begin{equation}
    \widetilde{O}_{M M'} (\mathbf p_1, \mathbf p_2, t) = a^6 \sum_{\mathbf{x},\mathbf{y}} e^{-i ( \mathbf{p}_1 \cdot \mathbf{x} + \mathbf{p}_2 \cdot \mathbf{y}) } O_{M M'} (x,y)
    \label{twobilinearmomprojection} \,,
\end{equation}
where $M M' \in \{ \pi\pi, K\pi\}$.

The suggestive operator labels in Eqs.~\eqref{rhokstarinterpolators}-\eqref{pionkaoninterpolators} are based on a quark-model construction for the corresponding hadrons. We emphasize that these operators cannot be directly mapped onto a set of finite-volume QCD states and ultimately, both $\widetilde O_V(\mathbf P, t)$ and $\widetilde{O}_{M M'} (\mathbf p_1, \mathbf p_2, t)$ will overlap all finite-volume states with the same isospin, strangeness, charge-conjugation, total momentum, and (for $\mathbf P = 0$) parity. It has, however, been empirically observed, that the relative overlap of a given operator to each finite-volume state, varies significantly between different operators. As is discussed in Sec.~\ref{sec:gevp}, this fact can be used to construct and solve a generalized eigenvalue problem, in order to reliably extract the finite-volume energies of interest.

A final important aspect of this construction is that, in addition to the quantum numbers listed above, the operators $\widetilde O_V(\mathbf P, t)$ and $\widetilde{O}_{M M'} (\mathbf p_1, \mathbf p_2, t)$ also carry angular momentum. In the finite-volume calculation, this label is converted to a finite-volume irreducible representation (irrep), drawn from a set of possibilities depending on the value of $\mathbf P$. This is discussed in more detail in Sec.~\ref{sec:opprojection}.

We turn now to a method for efficiently constructing correlation functions using the operators introduced here. To reduce the clutter of notation, in the following sections, we denote momentum-projected operators by $O$ instead of $\widetilde{O}$.

\subsection{Distillation}
\label{sec:distillation}

A significant computational challenge of this project is to obtain all the two-point correlation functions from the interpolators in Sec.~\ref{sec:operatorprojection},
\begin{align}
\begin{split}
    & \langle O_M(\mathbf P, t') \, O_M(\mathbf P, t)^\dagger \rangle \,, \\
    & \langle O_{MM'}(\mathbf p_1, \mathbf p_2, t') \, O_V(\mathbf P, t)^\dagger \rangle \,, \\ 
    & \langle O_{MM'}(\mathbf p_1, \mathbf p_2, t') \, O_{MM'}(\mathbf p_1', \mathbf p_2', t)^\dagger \rangle \,, 
\end{split}
\end{align}
for several momentum combinations later specified. For the correlation functions involving multi-hadron interpolators, propagators from all lattice sites to all other lattice sites have to be computed.
One strategy and an established tool in hadron spectroscopy to cost-efficiently achieve this is the so-called distillation method~\citep{Peardon:2009gh,Morningstar:2011ka}, which is used in this work. 
Distillation allows us to readily construct all operator combinations needed to build the required matrix of correlation functions. This comprises a straightforward way to compute correlation functions with operators on arbitrary time slices and with any lattice momentum needed.
One starts with the covariant three-dimensional Laplacian operator
\begin{align}
\begin{split}
    & - \mathbf \nabla_{ab}^2(\mathbf{x},\mathbf{y};t) \equiv 6 \delta_{\mathbf{x},\mathbf{y}} \delta_{ab} \\ 
    & \qquad - \sum_{j=1}^3 (U^{ab}_j(\mathbf{x},t)\delta_{\mathbf{x}+a\hat{\textbf j},\mathbf{y}} + U^{ba}_j(\mathbf{y},t)^{*}\delta_{\mathbf{x}-a\hat{\textbf j},\mathbf{y}} )
    \label{laplacian}
\end{split}
\end{align}
built from stout-smeared~\citep{Morningstar:2003gk} gauge links $U_j(\mathbf{x},t)$ connecting the spatial lattice sites $\mathbf{x}$ and $\mathbf{x}+a\hat{\textbf j}$, where $\hat{\textbf j}$ is the unit vector on the $j$th spatial direction and $a$ the lattice spacing. The color indices $a,b \in \{1,2,3\}$ will be absorbed into the matrix notation in the remainder of this section. The corresponding eigenvalue problem on each time slice is written as
\begin{equation}
    - \sum_{\mathbf{y}} \mathbf\nabla^2(\mathbf{x},\mathbf{y};t) v_k(\mathbf{y},t) = \xi_k(t) v_k(\mathbf{x},t)  \,,
\end{equation}
where the eigenvalues are ordered by magnitude $\xi_1(t) < \xi_2(t) < \ldots < \xi_{\Nvec}(t)$. The first $\Nvec$ eigenvectors $v_k(\mathbf{x},t)$ span the low-mode subspace of $-\mathbf \nabla^2$. The distillation-smearing kernel is the projector into distillation space and is defined via \citep{Peardon:2009gh}
\begin{equation}
    \square(\mathbf x, \mathbf y; t) \equiv \sum_{k=1}^{\Nvec} v_k(\mathbf x, t) v_k(\mathbf y, t)^\dagger = V(\mathbf x, t)V(\mathbf y, t)^\dagger \,,
    \label{dist-kernel}
\end{equation}
where the eigenvectors $v_k$ are organized as the columns of the ${N_c N_T N_L^3 \times \Nvec}$ rectangular matrix ${V \equiv \left( v_1 \ v_2 \ \ldots \ v_{\Nvec} \right)}$. Here, $N_c=3$, $N_L = L/a$ and $N_T = T/a$ are the number of colors, the spatial and temporal extensions in lattice units, respectively. We also use dilution projectors 
\begin{equation}
    P^{[d]}(t,\alpha,k)=\delta_{t,d_t} \delta_{\alpha,d_\alpha} \delta_{k,d_k}
\end{equation}
with a compound dilution index $[d]=[d_t,d_\alpha,d_k]$ projecting out time slice $t \in [0,N_T)$, spin index $\alpha \in [0,4)$ and Laplacian eigenvector $k \in [1,\Nvec]$. The Kronecker-delta symbols project out exactly one index-triple $(t,\alpha,k)$ per dilution index $[d]$, leading to what we call \textit{exact distillation}. More general dilution projectors are needed if one were to employ stochastic distillation~\citep{Morningstar:2003gk}.

Applying the smearing matrix $\square$ to both quark fields in a quark propagator $D^{-1}\equiv \langle q(x)\bar{q}(y) \rangle$, one obtains
\begin{align}
\begin{split}
    \left[ \square D^{-1} \square \right](x;y)&= \left[ V V^\dagger D^{-1} V V^\dagger \right](x;y)  \\
    &= \sum_{[d]} \left[ V V^\dagger D^{-1} (V P^{[d]})(P^{[d]} V^\dagger) \right](x;y)  \\
    &= \sum_{[d]} \left[ V \tau^{[d]} (P^{[d]} V^\dagger) \right](x;y)  \,,
    \label{dist-prop}
\end{split}
\end{align}
where we identify the so-called \textit{perambulator}
\begin{equation}
    \tau^{[d]}_{(f)}(t) \equiv a^6 \sum_{\mathbf x', \mathbf y'} V(\mathbf x', t)^\dagger D^{-1}_{(f)} (\mathbf x', t ; \mathbf y', d_t) [V(\mathbf y', d_t) P^{[d]}]  \,,
\end{equation}
for the propagation of a quark of flavor $f$. Note that $P^{[d]}$ has projected out a single time slice from $t'=d_t$ belonging to the corresponding dilution component. We can factorize Eq.~\eqref{dist-prop} into a \textit{source vector}
\begin{equation}
    \varrho_{\alpha}^{[d]}(\mathbf x, t) \equiv \sum_{k=1}^{\Nvec} v_k (\mathbf x, t) P^{[d]} (t, \alpha, k)  \,,
\end{equation}
and a \textit{sink vector}
\begin{equation}
    \varphi_{\alpha(f)}^{[d]}(\mathbf x, t) = \sum_{k=1}^{\Nvec} v_k(\mathbf x, t) \tau_{\alpha k (f)}^{[d]}(t)  \, .
\end{equation}
For instance, using these objects, the two-point correlation function from the $\bar d \gamma_5 u$ interpolator in Eq.~\eqref{pionkaoninterpolators} can be written as
\begin{align}
\begin{split}
    & \langle O_{\pi^+}(\mathbf P, t') O_{\pi^+}(\mathbf P, t)^\dagger \rangle_F = a^6 \sum_{\mathbf{x},\mathbf{y}} e^{-i \mathbf P \cdot (\mathbf{x'} - \mathbf{x}) } \\
    & \qquad\qquad \times \sum_{d_1, d_2} \tr \big[ \varrho^{[d_1]}(\mathbf{x}',t')^{\dagger} \gamma_5  \varphi_{(u)}^{[d_2]}(\mathbf{x}',t') \big] \\
    & \qquad\qquad\qquad \times \tr \big[ \varrho^{[d_2]}(\mathbf{x},t)^{\dagger} \gamma_5 \varphi_{(d)}^{[d_1]}(\mathbf{x},t) \big]  \\
    & \! = \sum_{d_1, d_2} \big( \dmf_{\gamma_5}^{(u)} \big)^{[d_1] [d_2]} (\mathbf P, t') \big( \dmf_{\gamma_5}^{(d)} \big)^{[d_2] [d_1]} (-\mathbf P, t)  \,,
    \label{simplecorrelatorcontractiondmf}
\end{split}
\end{align}
where the $4 N_T \Nvec \times 4 N_T \Nvec$ \textit{meson field} matrix is defined for each $\Gamma, \mathbf p, t$ and $f$ as
\begin{align}
\begin{split}
    &\big( \dmf^{(f)}_{\Gamma} \big)^{[d_1] [d_2]} (\mathbf p, t) \\
    &\qquad = a^3 \sum_{\mathbf{x}} e^{-i \mathbf{p} \cdot \mathbf{x}} \tr \big[ \varrho^{[d_1]} (\mathbf x, t)^{\dagger} \, \Gamma \, \varphi^{[d_2]}_{(f)}(\mathbf x, t) \big]  \,,
    \label{mesonfieldrhophi}
\end{split}
\end{align}
with traces over color and spin~\citep{Morningstar2013}. The subscript $F$ on the angled brackets means the expectation value only with respect to the fermionic path integral in a fixed gauge configuration. A similar factorization can be utilized for more complicated correlation functions involving the distillation-smeared versions of the operators in Eqs.~\eqref{rhokstarinterpolators},\eqref{pipikpiinterpolators},\eqref{pionkaoninterpolators}. In our implementation, we generate meson fields with momentum projection up to $0 \leq \mathbf{p}^2 \leq 4 \left( \frac{2\pi}{L} \right)^2$ for the matrices $\Gamma \in \{ \gamma_5, \gamma_x, \gamma_y, \gamma_z \}$ and quark flavors $f \in \{ l, s\}$, where $s$ stands for the strange quark and $l$ stands for the degenerate light quarks in the isospin-symmetric limit, due to our computational setup detailed in Sec.~\ref{sec:ensemble}. The combination of Wick contractions leading to correlators with well-defined isospin quantum numbers are diagrammatically shown in Appendix~\ref{apx:wick} in terms of meson fields.

In this work, we use the \textit{Grid} and \textit{Hadrons} open-source libraries~\citep{Boyle2015, Hadrons2023}. In particular, dedicated \textit{Hadrons} modules were developed for computing the perambulators and meson fields needed in this and other works~\citep{Hadrons2023, Boyle2019}, including an ongoing study of hadronic $D \to K\pi$ decays at the $SU(3)$-flavor symmetric point~\citep{Joswig:2022ctr}.

\subsection{Operator Projection}
\label{sec:opprojection}

In infinite-volume QCD, creation and annihilation operators of integer-spin states transform under $SO(3)$, and so have well-defined angular momentum quantum numbers. On the other hand, operators on a finite lattice are restricted to subgroups of $SO(3)$ corresponding to symmetry transformations of cuboid shapes. This implies that the interpolators defined in Sec.~\ref{sec:operatorprojection}, and thus the multi-particle states created by them, will transform according to reducible representations of such restricted symmetries. As a direct consequence, the spectral decomposition of the correlators computed from the operators~\eqref{rhokstarinterpolators},\eqref{pipikpiinterpolators},\eqref{pionkaoninterpolators} will contain potentially significant contributions from states belonging to different lattice irreducible representations (irreps), hindering the extraction of the associated finite-volume energies. 

\begin{table}[ht!]
    \centering
    \setlength{\tabcolsep}{0.5em}{\renewcommand{\arraystretch}{1.5}
    {
    \begin{tabular}{cc}
        \hline
        \multicolumn{1}{c|}{$\Lambda[\mathbf d]$} & Reference two-bilinear momenta                 \\\hline\hline
        \multicolumn{2}{c}{$K\pi$ $(I=1/2): [\mathbf d_K][\mathbf d_\pi]$}                                  \\\hline
        \multicolumn{1}{c|}{$T_{1u}[000]$} & [001][00-1], [110][-1-10], [111][-1-1-1], [002][00-2] \\\hline
        \multicolumn{1}{c|}{$E[001]$}      & [101][-100], [-100][101], [1-11][-110], [-110][1-11]  \\\hline
        \multicolumn{1}{c|}{$B_1[110]$}    & [10-1][011], [111][00-1], [00-1][111]                 \\\hline
        \multicolumn{1}{c|}{$B_2[110]$}    & [100][010], [011][10-1], [-110][200], [200][-110]     \\\hline
        \multicolumn{1}{c|}{$E[111]$}      & [101][010], [010][101], [1-11][020], [020][1-11]      \\\hline
        \multicolumn{1}{c|}{$E[002]$}      & [011][0-11], [1-11][-111]                             \\\hline\hline
        \multicolumn{2}{c}{$\pi\pi$ $(I=1): [\mathbf d_\pi][\mathbf d_\pi]$}                              \\\hline
        \multicolumn{1}{c|}{$T_{1u}[000]$} & [001][00-1], [110][-1-10], [111][-1-1-1], [002][00-2] \\\hline
        \multicolumn{1}{c|}{$E[001]$}      & [101][-100], [1-11][-110]                             \\\hline
        \multicolumn{1}{c|}{$B_1[110]$}    & [10-1][011], [111][00-1]                              \\\hline
        \multicolumn{1}{c|}{$B_2[110]$}    & [100][010], [011][10-1], [-110][200]                  \\\hline
        \multicolumn{1}{c|}{$E[111]$}      & [101][010], [1-11][020]                               \\\hline
        \multicolumn{1}{c|}{$E[002]$}      & [011][0-11], [1-11][-111]                             \\\hline
        \multicolumn{1}{c|}{$A_1[001]$}    & [000][001], [-100][101], [-1-10][111], [00-1][002]    \\\hline
        \multicolumn{1}{c|}{$A_1[110]$}    & [000][110], [111][00-1], [200][-110]                  \\\hline
        \multicolumn{1}{c|}{$A_1[111]$}    & [111][000], [110][001], [11-1][002]                   \\\hline
        \multicolumn{1}{c|}{$A_1[002]$}    & [000][002], [001][001]                                \\\hline
    \end{tabular}
    }}
    \caption{Reference total momentum $\mathbf d$ and individual reference momenta assignments $[\mathbf d_1][\mathbf d_2]$ of two-bilinear operators in the corresponding lattice irreps they were projected into, using the Schönflies notation~\citep{Elliot1979a}. We only use irreps whose leading subduction comes from the $\ell=1$ partial wave and do not mix with even $\ell$ in the case of $K\pi$. Additionally, we only use irreps in the little groups with total momenta up to $\mathbf d^2 = 4$.}
    \label{tab:irrepmomenta}
\end{table}
For this reason, it is crucial to project interpolating operators into the irreps $\Lambda[\mathbf{d}]$ of the lattice symmetries in the various possible moving frames~\citep{Thomas2012, Morningstar2013, Detmold2024b}, where $\mathbf d$ denotes the dimensionless total momentum according to
\begin{equation}
    \mathbf d \equiv \frac{L}{2\pi} \mathbf P \,.
\end{equation}
We use a similar notation for the individual momenta of the two-bilinears \eqref{twobilinearmomprojection}, obeying total momentum conservation
\begin{equation}
    \mathbf d_1 + \mathbf d_2 = \mathbf d  \,,
    \label{twobilinearmomenta}
\end{equation}
where $\mathbf d_1, \mathbf d_2$ are either $\mathbf d_{K}$ or $\mathbf d_{\pi}$. At rest, the relevant symmetry is given by the octahedral group $\mathscr{O}_h$, composed of the $48$ transformations leaving a cube invariant, including spatial inversions~\citep{Elliot1979a}. In moving frames ($\mathbf{d}^2 > 0$), the appropriate symmetries can be identified with the subgroups of $\mathscr{O}_h$ under which a given overall momentum $\mathbf d$ is invariant, also known as \textit{little groups} of $\mathbf d$~\citep{Moore:2005dw}. We list the relevant little groups for the operators used in this work in Table~\ref{tab:irrepmomenta}, whose associated total momentum is considered in the range
\begin{equation}
    0 \leq \mathbf d^2 \leq 4  \, .
\end{equation}
To de-clutter notation, we will often keep the total momentum implicit and use $\Lambda$ to denote both irrep and total momentum, unless explicitly needed.

We implement the projections of operators depending on $n_{\mathbf p}=1$ or $n_{\mathbf p}=2$ momenta through the group-theoretical formula~\citep{Elliot1979a, Prelovsek2017} 
\begin{align}
\begin{split}
    O^{\Lambda r}(\mathbf p_1 &, \ldots, \mathbf p_{n_{\mathbf p}}, t) =  \\ 
    &   \sum_{R \in G} B^{\Lambda r}(R) \, \hat R O(\mathbf p_1, \ldots, \mathbf p_{n_{\mathbf p}}, t) \hat R^{-1} \,,
    \label{generalprojector_row}
\end{split}
\end{align}
where the transformation $R$ of the group $G$ is performed at the operator level via $\hat R$, and $r$ is a certain row of the representation matrix of $\Lambda$. The coefficients $B^{\Lambda r}$ encode a number of choices when selecting the representation matrices and the reference momenta $\mathbf p_1, \mathbf p_2$ (see Table~\ref{tab:irrepmomenta}), being only required that the resulting operators are nonzero and linearly independent~\citep{Foley2011,Gockeler2012a}. In the case of $n_{\mathbf p}=1$, acting on the single-bilinear operators of Eq.~\eqref{singlebilinearmomproj} leads to operators $O^{\Lambda r}_{M}$ with definite row and irrep quantum numbers. For $n_{\mathbf p}=2$, the Clebsch-Gordan coefficients for combining the $K$ and $\pi$-like $O_M$ interpolators into the lattice-projected operators $O^{\Lambda r}_{MM'}$ are inferred from the use of the projection formula above on the operators of Eq.~\eqref{twobilinearmomprojection}.

Using the lattice-projected interpolators, we form the \textit{interpolator bases}
\begin{align}
\begin{split}
    & \big\{\,  O_X^{\mathcal{Q}}(t) \,\big\} \equiv \Big\{ \, O^{\Lambda r}_V(\mathbf P, t)\,, \  O^{\Lambda r}_{MM'}(\mathbf p_1^{[1]}, \mathbf p_2^{[1]}, t) \,,\ \\
    & O^{\Lambda r}_{MM'}(\mathbf p_1^{[2]}, \mathbf p_2^{[2]}, t) \,,\ \ldots, \ O^{\Lambda r}_{MM'}(\mathbf p_1^{[n_\mathrm{op}^{\mathcal{Q}}-1]}, \mathbf p_2^{[n_\mathrm{op}^{\mathcal{Q}}-1]}, t) \,\Big\}  \,,
    \label{operatorbasis}
\end{split}
\end{align}
where $\mathcal{Q}$ represents all flavor and irrep quantum numbers listed in Table~\ref{tab:irrepmomenta}, and ${V \in \{ \rho^+, K^{*+} \}}$ and ${MM' \in \{ \pi\pi, K\pi \}}$. The indices in square brackets are used here only to count the number of two-bilinear operators. The compound index ${X = 1,\, \ldots,\, n_\mathrm{op}^{\mathcal{Q}}}$ collectively labels the bilinear structure ($V$ or $MM'$) and the respective momenta assignments in Table~\ref{tab:irrepmomenta}.

\subsection{Generalized Eigenvalue Problem}
\label{sec:gevp}

Using the meson fields \eqref{mesonfieldrhophi}, we compute all correlators from the combination of interpolators in the operator bases \eqref{operatorbasis} with given quantum numbers $\mathcal{Q}$. We perform this at all $N_T$ source times, which are then translated to the temporal origin and averaged. To further improve the signal-to-noise ratio, the correlators projected to equivalent total momenta under little group symmetries are identified and averaged~\citep{Morningstar2013}. When an irrep is not one-dimensional, we also average over irrep rows and drop their label $r$ from the operator basis in Eq.~\eqref{operatorbasis}.

\begin{figure*}
    \centering
    \includegraphics[width=17.2cm]{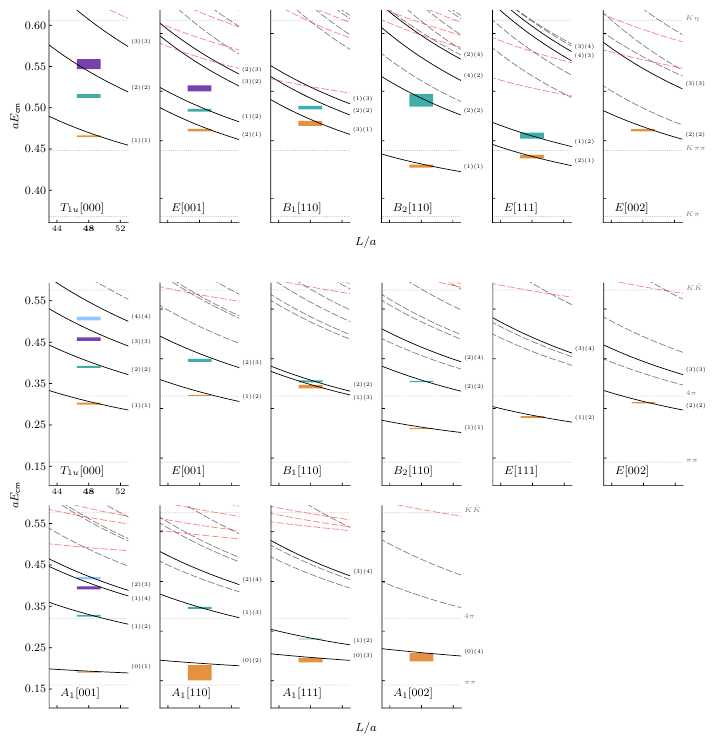}
    \caption{Non-interacting two-particle energies given by Eq.~\eqref{nonintCMformula} with the momenta assignments of the $K\pi$ (top) and $\pi\pi$ (bottom) two-bilinear operators that are included in the GEVP (full black lines labeled by $(\mathbf d_{K}^2) (\mathbf d_{\pi}^2)$ and $(\mathbf d_{\pi}^2) (\mathbf d_{\pi}^2)$, respectively). They are respectively distributed over $6$ and $10$ irreps on the $5$ momentum frames considered, where the odd-even wave mixing irreps were not considered in the $K \pi$ analysis. The horizontal axes show the variation of the free levels with the box size $L$ at constant lattice spacing $a$, where $a$ and $L/a=48$ correspond to the ensemble used in this work. The free energies corresponding to two-bilinear operators not included (where at least one meson has momentum $|\mathbf d|^2>4$) are also shown (dashed gray lines). The red dashed lines represent the free low-lying $K\pi\pi$ and $\pi\pi\pi\pi$ states in the corresponding irreps. We also show the model-averaged finite-volume energies extracted in this work (colored boxes), as described in Sec.~\ref{sec:spectrumdetermination}. The same energy levels overlaid to the lattice data can be seen in Appendix~\ref{apx:gevpeffspectra}. Any finite-volume energy above a grey or red dashed line does not enter the phase shift determination (Sec.~\ref{sec:phaseshiftdetermination}) and is also not shown in this figure. The relevant scattering thresholds are also shown for convenience (faint dots).}
    \label{fig:nonint_modelavg_spectrum}
\end{figure*}

We can finally build the $n_\mathrm{op}^{\mathcal{Q}} \times n_\mathrm{op}^{\mathcal{Q}}$ matrices of correlation functions
\begin{equation}
    C^{\mathcal{Q}}_{XZ}(t) \equiv \langle O^{\mathcal{Q}}_{X}(t) O^{\mathcal{Q}}_{Z}(0)^{\dagger} \rangle
    \label{corrmatrix}
\end{equation}
from the bases Eq.~\eqref{operatorbasis}, and solve the generalized eigenvalue problem (GEVP)~\citep{Luscher1990a,Blossier:2009kd,Fischer2020a}
\begin{align}
    C^{\mathcal{Q}}(t) u^{\mathcal{Q}}_n(t)  = \lambda^{\mathcal{Q}}_n(t) C^{\mathcal{Q}}(t_0) u^{\mathcal{Q}}_n(t)  \,,
    \label{gevpequation}
\end{align}
at fixed $t_0$, resulting in $n_\mathrm{op}^{\mathcal{Q}}$ generalized eigenvalues $\lambda^{\mathcal{Q}}_n(t)$ and eigenvectors $u^{\mathcal{Q}}_n(t)$, where their $t_0$ dependence is left implicit. The time slice $t_0$ is chosen not too large in order to not feed excessive noise into the GEVP solution. A specific choice of $t_0$ will be made in Sec.~\ref{sec:analysis}. For a fixed $t_0$, it can be shown that the GEVP eigenvalues have the asymptotic form~\citep{Luscher1990a, Blossier:2009kd}
\begin{align}
    \lambda^{\mathcal Q}_n(t) =  Z_n^{\mathcal Q} \exp \! \left ( -E_n^{\mathcal Q} t \, \right ) \left [1 + \mathcal O(e^{- \Delta^{\mathcal Q}_n t}) \right ]  \,,
    \label{gevpequation-eigenvalue}
\end{align}
where $E_n^{\mathcal Q}$ denotes a finite-volume energy level in the spectrum and $\Delta^{\mathcal Q}_n = \min_{m \neq n} | E^{\mathcal{Q}}_n - E^{\mathcal{Q}}_m | > 0$
encodes the residual excited state contamination. A suitable analysis of the eigenvalues $\lambda^{\mathcal{Q}}_n(t)$, which we often refer to as correlators, allows the extraction of the relevant finite-volume energies in all moving frames and irreps considered. From now on, each irrep and channel labeled by $\mathcal{Q}$ and state $n$ will be collectively labeled by $(i)=\{\mathcal{Q}, n \}$, such that $\lambda^{\mathcal{Q}}_n \equiv \lambda^{(i)}$.

We stress that the largest momentum we give to each individual interpolator bilinear is of magnitude $\mathbf{d}_h^2=4$, where $\mathbf d_h \in \{ \mathbf d_K, \mathbf d_\pi \}$ and $h$ is an index over all the particles considered. As illustrated in Fig.~\ref{fig:nonint_modelavg_spectrum}, we have some interpolators whose momentum assignments correspond to non-interacting energies,
\begin{equation}
    E^\mathrm{free}_\cm(L)^2 = \left (\sum_h \sqrt{m_h^2 + ( 2\pi/L )^2 \, \mathbf d_h^2 } \right )^{\!\!2} - \mathbf P^2 \,,
    \label{nonintCMformula}
\end{equation}
above other ones including a bilinear with ${\mathbf{d}_h^2 > 4}$, due to certain combinations of momenta. One such example is in the $B_2[110]$ irrep in Fig.~\ref{fig:nonint_modelavg_spectrum}, where the non-interacting energy formed by the momenta ${\mathbf{d}_K=[-100]}$ and ${\mathbf{d}_{\pi}=[210]}$ is lower than the one from ${\mathbf{d}_K=[-110]}$ and ${\mathbf{d}_{\pi}=[200]}$ (lowest gray dashed line). Additionally, there can also be three- or four-particle non-interacting energies that lie lower in the spectrum than the ones corresponding to the two-bilinear operators we employ. Our criterion to determine the $n^\mathrm{lev}$ energies going into the phase shift determination in Sec.~\ref{sec:phaseshiftdetermination} is that we exclude all the ones lying \textit{at or above} the lowest non-interacting energy whose operator we do not employ. In this way, we avoid using levels with potentially significant overlaps to such ``missing" operators. We also do not use the levels on top or above the $K \eta$ and $K\bar K$ thresholds, respectively in the $K\pi$ and $\pi\pi$ channels.

\bigskip
\subsection{Finite-volume Method}
\label{sec:phaseshift}

In the infinite-volume continuum theory, the scattering matrix or $S$-matrix is diagonal in the angular momentum basis. In addition, it is a unitary matrix (due to the unitarity of the theory), implying that partial-wave components for an elastic two-to-two process can be written in terms of real phases $\delta_\ell(E_\cm)$ as
\begin{equation}
    S_{\ell m, \ell' m'}(E_\cm) = \delta_{\ell \ell'} \delta_{mm'} e^{2 i \delta_\ell(E_\cm)}  \,,
\end{equation}
where $\ell,m$ label the incoming angular momenta and $\ell', m'$ the outgoing, and $E_\cm$ is the total energy in the c.m. frame.

Both the $\pi\pi$ and $K\pi$ channels admit a description of this form and in this work, we are specifically interested in the $\ell=1$ phase shift for each. As was mentioned in the introduction, these channels are only strictly elastic for energies below the nearest three- or four-hadron threshold. For energies above these thresholds, $\delta_1(E_\cm)$ acquires a small imaginary part. However, the effect of this on the extracted resonance parameters is expected to be much smaller than other sources of systematic uncertainty, as we detail in Sec.~\ref{sec:results}, and so we take the elastic approximation throughout.

In the Lüscher method, the phase shift can be calculated using a quantization condition~\citep{Luscher:1986pf, Kim:2005gf, Hansen:2012tf} which constrains the scattering amplitude evaluated at energies from the finite-volume spectrum. In the case that only a single partial wave is relevant, the quantization condition can be written in the pseudophase form~\citep{Alexandrou:2017mpi,Rendon:2020rtw,Leskovec:2012gb}
\begin{equation}
    \delta_1 \big( E_\cm^{(i)}(L) \big) = n \pi - \phi^{[\mathbf d, \Lambda]} \left(E_\cm^{(i)}(L), L, m_1, m_2 \right)  \,,
    \label{pseudophaseqcform}
\end{equation}
for $n \in \mathbb Z$. Here $\phi$ is a known geometric function, sometimes called the pseudophase. We emphasize here that the two sides are equal only when evaluated at $E_\cm^{(i)}(L) = \sqrt{E^{(i)}(L) ^2 - \mathbf P^2}$ where $E^{(i)}(L)$ is a finite-volume energy.

The pseudophase $\phi$ is defined in terms of the geometrical function $F$ (and the latter is defined in terms of the generalized zeta function $\mathcal{Z}$) as follows:
\begin{align}
    & \cot \phi^{[\mathbf d, \Lambda]} \left(E_\cm, L, m_1, m_2 \right) = \nonumber\\
    & \qquad\qquad \mathbb P^{[\mathbf d,\Lambda]}_{m}  \, i F_{1 m, 1 m'}^{\mathbf d}(q, L, m_1, m_2) \, \mathbb P^{[\mathbf d, \Lambda]*}_{m'} \,, \label{phidef}\\
    & F_{\ell m, \ell' m'}^{\mathbf d}(q, L, m_1, m_2) = \nonumber\\
    & \qquad\qquad \frac{(-1)^\ell}{\gamma \pi^{3/2}} \sum_{j=|\ell-\ell'|}^{\ell+\ell'} \sum_{s=-j}^j \frac{i^j}{q^{j+1}} \mathcal{Z}_{js}^{\mathbf d}(q^{2}) \, C_{\ell m,js,\ell'm'} \,.
    \label{geometricalfunction}
\end{align}
To unpack these relations we first note the change of coordinates from $E_\cm$ to $q$ in Eq.~\eqref{phidef}. The relation between the two follows from $q = p_\cm L/(2 \pi)$ and
\begin{equation}
    \label{CMenergies}
    E_\cm  = \sqrt{p_\cm^2 + m_1^2} + \sqrt{p_\cm^2 + m_2^2} \, .
\end{equation}
This enters simply because it is conventional to express $F$ in terms of $q$ while we find it most convenient to express $\delta_1$ and thus $\phi$ in terms of $E_\cm$. In Eq.~\eqref{phidef} we have additionally included projection operators $\mathbb P^{[\mathbf d,\Lambda]}_{m}$ which specify a definite finite-volume irrep $\Lambda$ taken from those listed in Table~\ref{tab:irrepmomenta}. The explicit form of these projections can be found in Refs.~\citep{Alexandrou:2017mpi, Rendon:2020rtw}. Finally, on the right-hand side of Eq.~\eqref{geometricalfunction} we have introduced $C_{\ell m,js,\ell'm'}$, $3j$-Wigner symbols, and $\mathcal{Z}_{js}^{\mathbf d}(q^{2})$, the generalized zeta functions. Both are defined in Refs.~\citep{Luscher:1986pf,Rummukainen:1995vs}.

In the finite-volume system, parity is not a good quantum number for nonzero total momentum, \ie $\mathbf P \neq 0$~\citep{Leskovec:2012gb,Wilson:2014cna}. This is particularly important for two-particle systems with non-degenerate masses, such as in the $K\pi$ case, since it can lead to mixing between even and odd partial waves. In particular, even and odd continuum partial waves both contribute to all $A_1$ irreps with nonzero total spatial momentum. At the level of the quantization condition, this means that the formula above would have to account for $\delta_0$, which is known to be non-negligible and phenomenologically important~\citep{Wilson:2015dqa,Rendon:2020rtw,Brett:2018jqw}. Therefore, the form of Eq.~\eqref{pseudophaseqcform} is valid only for the irreps that do not simultaneously subduce into $S$ and $P$ waves, \cf Table~\ref{tab:irrepmomenta}. Even though it is known how to treat such types of quantization conditions, we restrict ourselves to the case of a single partial wave extraction and reserve the $S$-wave physics for future work.

By applying Eq.~\eqref{pseudophaseqcform} in multiple moving frames, one can constrain the $P$-wave elastic phase shift $\delta_1$ as described in Sec.~\ref{sec:phaseshiftdetermination}. In particular, resonance parameters can be extracted by using different phase-shift models. Ultimately, such models are embedded back into the scattering amplitude, related to the phase shift according to
\begin{align}
    \tmat{\ell}(E_\cm)  =  \frac{1}{\cot \delta_\ell(E_\cm) - i} \, ,
   \label{tmatrixdef}
\end{align}
where $ S_{\ell m, \ell' m'} = \delta_{\ell \ell'} \delta_{mm'}  \left( 1 + i \tmat{\ell} \right)$.
This is then analytically continued into the complex energy plane, giving us information on the resonance pole parameters (see Sec.~\eqref{sec:poleposition}).

\bigskip
\section{Computational Setup}
\label{sec:setup}

\subsection{Ensemble}
\label{sec:ensemble}

We perform our calculation on a $2+1$ flavor RBC/UKQCD Möbius domain-wall fermion (DWF) ensemble~\citep{RBC:2014ntl} employing the Iwasaki gauge action. The inverse lattice spacing is $a^{-1}=\valueainverse$ and the volume is $(L/a)^3 \times (T/a)=48^3 \times 96$, with the DWF-specific fifth dimension having length $L_s=24$ and the domain wall height $aM_5 = 1.8$. The bare quark masses on this ensemble are $am_l=0.00078$ and $am_s=0.0362$. The valence light-quark action employed in this work uses the zMöbius approximation of the domain-wall Dirac operator which reduces the effective fifth dimension length to $L_s=10$~\citep{McGlynn2015}. The resulting valence pion and kaon masses, as determined in this work, are $m_\pi=\valueourpionmass$ and $m_K=\valueourkaonmass$, respectively. The use of the non-unitary action for the valence sector introduces a systematic bias that must, however, vanish in the continuum limit. As we have seen in earlier works performed on the same ensemble~\citep{RBC:2022ddw,Boyle:2022lsi,AndrewThesis2022}, this leads to a difference in the pion mass of $\sim$1\% and we neglect this effect in this work\footnote{This effect is of similar magnitude as the one expected from isospin breaking corrections, which we neglect as well.}.

We compute correlation functions on $90$ gauge configurations (separated by $20$ Monte-Carlo time steps each) involving quark propagators on all $96$ source-time displacements to increase statistics. We average measurements from different source-time displacements and equivalent irrep rows and total momentum, resulting in one estimator per configuration. The light-quark Dirac operator inversions are done using a red-black preconditioned conjugate gradient algorithm with a mixed-precision approach within the \textit{Grid} and \textit{Hadrons} open-source libraries~\citep{Boyle2015, Hadrons2023}. In the light-quark sector, we make use of deflation with the low-lying $2000$ eigenvectors of the Dirac operator~\citep{Luscher2010c}, employed both on CPU and GPU architectures. The Dirac operator eigenvectors were computed with a Lanczos algorithm under Chebyshev preconditioning. The eigenvectors of the covariant three-dimensional-Laplacian operator required in distillation, Eq.~\eqref{laplacian}, were computed in a similar way.

\begin{figure*}[t!]
    \centering
    \includegraphics[width=\linewidth]{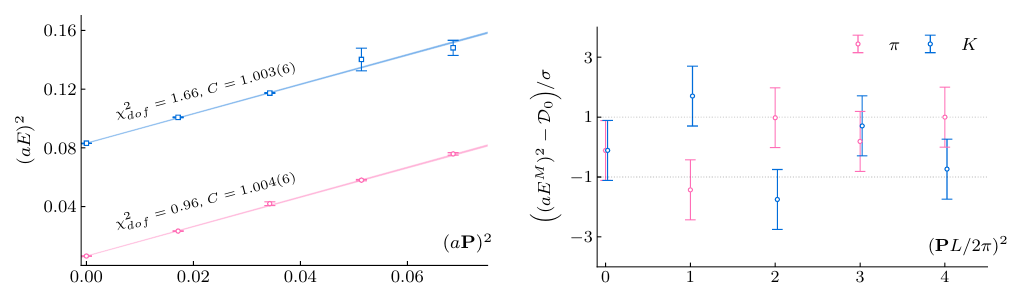}
    \caption{\textit{Left}: Statistical error bands from the fit of the model \eqref{adjustablecontdispersionmodel} to the pseudoscalar moving frame energies $a E^M$ obtained from correlators involving $\bar q \gamma_5 q'$-type interpolators only. Their reduced chi-squares are shown next to the respective bands, together with the optimal slope parameter $C$. \textit{Right}: Difference between data and fit results normalized by the standard deviation $\sigma$ of the data $(a E^M)^2$.}
    \label{pionkaondispersion}
\end{figure*}

\subsection{Dispersion Relation}
\label{sec:dispersion}

In this section, we confirm that the continuum relativistic dispersion relation
\begin{equation}
    E^2 = m^2 +  \mathbf{P}^2  \,,
    \label{continuumdispersionrelation}
\end{equation}
holds to a sufficient approximation for single pions and kaons such that the continuum finite-volume formalism can be sensibly applied. Using the interpolators defined in \eqref{pionkaoninterpolators}, we obtain the correlators $\langle O_M(\mathbf P, t) \, O_M(\mathbf P, 0)^\dagger \rangle$, $M \in \{ \pi^+, K^+ \}$, after having averaged over all available $N_T=96$ source time slices and rotationally-equivalent total momenta $\mathbf P$ in a given moving frame. We use the `single-exp'
\begin{align}
    \mathcal{C}_M(t) = Z^M \exp(- E^M t) \,,
\end{align}
and the `single-cosh'
\begin{align}
    \mathcal{C}_M(t) = Z^M \left[ \exp(- E^M t) + \exp(- E^M (T-t)) \right] \,,
\end{align}
models to fit directly to such correlator data. We give the energy argument a momentum label $E^M(\textbf P)$ to track the fact that its value depends on the momentum of the correlator in the fit. This model parameter represents the low-lying energy present in the spectral decomposition of the bilinear correlators, while $Z^M(\textbf P)$ is the corresponding overlap factor.

We perform \textit{all correlator fits} including at least four consecutive time slices and imposing that the largest time slice considered is in a region where the signal-to-noise is still at a reasonable level. For the purpose of checking the dispersion relation through the models below, which are not directly used further in the analysis, we adopt a simpler approach than that of Sec.~\ref{sec:spectrumdetermination}. Here, after ensuring consistency between different choices of correlator model, we pick one list of representative fits, namely the single-cosh results with the minimum value for the Akaike information criterion~\citep{Akaike1974} ($\AIC$, see Sec.~\ref{sec:spectrumdetermination}) for each $\mathbf P$, to obtain moving-pion and kaon energies. We then fit the continuum-like model with free parameters $A,C$,
\begin{equation}
    \mathcal D_{0}(\mathbf{P}^2) = A + C (a \mathbf{P})^2  \,,
    \label{adjustablecontdispersionmodel}
\end{equation}
to such dataset $\left(a E^M\right)^2(\textbf P)$ of lattice-unit energies at total momenta ${\left(\frac{L}{2\pi}\mathbf P\right)^2 \in \{ 0,1,2,3,4 \} }$. As it is depicted in Fig.~\ref{pionkaondispersion}, the data is well described by \eqref{adjustablecontdispersionmodel}, yielding $C  = 1$ within one standard deviation. We also try a similar model with an $a^2$ correction to the usual continuum form
\begin{equation}
    \mathcal D_{a^2}(\mathbf{P}^2) = A + C (a \mathbf{P})^2 + F (a \mathbf{P})^4  \,,
    \label{adjustablecontdispersionmodelasqr}
\end{equation}
which again shows reasonable $\chi^2$ but yields the parameter $F$ consistent with zero within one standard deviation. This suggests we cannot resolve $\mathcal{O}(a^2)$ effects at the level of the dispersion relation of single pion and kaon states. 

Based on the consistency of the data with the dispersion relation models above, we use the exact relation \eqref{continuumdispersionrelation} to later boost finite-volume energies to the c.m. frame, \cf Eq.~\eqref{cmenergyresultdefinition}. Throughout the scattering analysis, we use pion and kaon masses determined from fits to correlators at zero total momentum, which have the highest statistical precision. We estimate their model-variation systematic uncertainty with the methods from Sec.~\ref{sec:modelaveraging}, including single-cosh and exponentials, and also two-cosh and exponentials. This yields $m_\pi=\valueourpionmass$ and $m_K=\valueourkaonmass$, whose central value is used throughout Eqs.~\eqref{nonintCMformula}-\eqref{CMenergies}. These are consistent with the values obtained from the parameter $A$ in Eqs.~\eqref{adjustablecontdispersionmodel},\eqref{adjustablecontdispersionmodelasqr}. We note that the finite-volume ground state differs from the infinite-volume mass by exponentially suppressed volume effects, which we neglect given the value of $m_\pi L \approx 4$ in our ensemble together with the fact that, at fixed $m_\pi L$, such effects are further suppressed as one decreases the physical value of the pion mass.

\subsection{Smearing Radius}
\label{sec:smearingradius}

It is important to correctly choose the number of distillation eigenvectors $\Nvec$, from which the smearing kernel Eq.~\eqref{dist-kernel} is built, given our physics goals and computation constraints. A larger value of $\Nvec$ leads to a narrower smearing profile, but also increases the cost of the computation: In exact distillation, employed in this work, the number of inversions of the Dirac matrix scales linearly with $\Nvec$.

The quark smearing present in distillation is used to improve the overlap of lattice interpolators to the low-lying spectrum of QCD. In that sense, as the number of distillation vectors $\Nvec$ is reduced, the smearing profile becomes broader, and the overlap to excited states reduces in a non-trivial way. However, when smearing too much, \ie choosing a value of $\Nvec$ that is too low, the overlaps to the desired low-lying excited states are also affected, resulting in a deterioration of the signal-to-noise ratio of the computed correlators. 

\begin{figure}[t!]
    \centering
    \includegraphics[width=8.4cm]{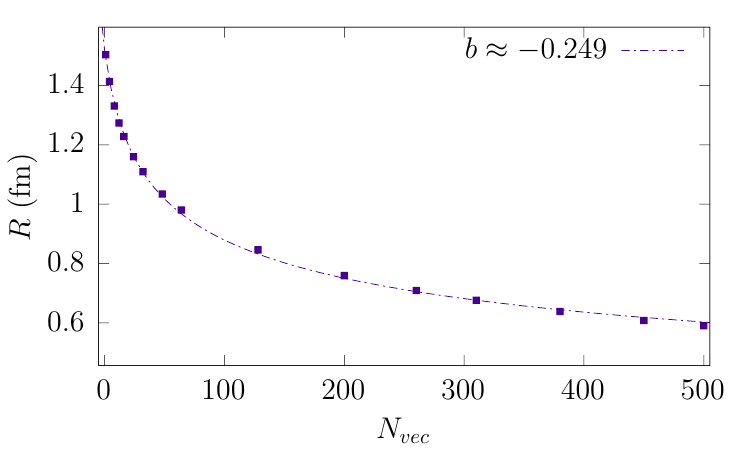}
    \caption{Smearing radius defined in Eq.~\eqref{smearingradius} as a function of $\Nvec$ (points). To guide the eye, we show a fit to a $a(\Nvec-c)^b$ model (dashed line)~\citep{Lachini2022}.}
    \label{fig:smearingradius}
\end{figure}

To assess the smearing effect of the $\square(\mathbf x, \mathbf y; t)$ (\cf Eq.~\eqref{dist-kernel}) on the spatial distribution of a point source, we compute~\citep{Peardon:2009gh}
\begin{equation}
    \label{spatialprofile}
    \Psi(\mathbf r) = \Big\langle \sum_{\mathbf x , t} \sqrt{ \tr \left[ \square(\mathbf x , \mathbf x + \mathbf r;t) \, \square(\mathbf x + \mathbf r , \mathbf x;t) \right] } \Big\rangle \,,
\end{equation}
where the trace is taken over the color indices of $\square$. In general, $\Psi$ also depends on the stout-smearing parameters $\rho_{\mathrm{stout}}, n_{\mathrm{stout}}$ used to smear the gauge-links that the three-dimensional Laplacian \eqref{laplacian} is constructed from~\citep{Morningstar:2003gk}.
We observed only a small dependence of $\Psi$ on different stout-smearing parameters and opted for ${n_{\mathrm{stout}}=3}$ iterations using a stout-smearing weight ${\rho_{\mathrm{stout}}=0.2}$ in the spatial directions~\citep{Peardon:2009gh}. As expected, as $\Nvec$ is increased the distribution approximates that of a point source. When working with hadron interpolators in momentum space, it is also clear that one needs a sufficiently localized spatial distribution to effectively project a source to definite total momentum.

We define a smearing radius $R$ as the $68.3\%$  percentile of the smearing profile around the origin, implicitly given by
\begin{equation}
    \label{smearingradius}
  \int_0^{R} \Psi(r) d r = 0.683 \int_0^{L/2} \Psi(r) d r  \,,
\end{equation}
for each $\Nvec$, which is straightforward to be numerically evaluated. We observe its dependence on $\Nvec$ to be approximately described by a power law with exponent $b \approx -0.25$ (Fig. \ref{fig:smearingradius}). We also observe an inflection point around $\Nvec \sim 60$ which, in other words, means that the smearing radius reduces less drastically above this value. 

After a signal-to-noise analysis of correlators formed from the operators $O_V$ in Eq.~\eqref{singlebilinearmomproj} within the range ${20 \le \Nvec \le 160}$, we adopted $\Nvec = 64$ to extract the few low-lying energies on moving frames up to $\mathbf d^2 = 4$~\citep{Lachini2022}. At this level of smearing, our cost-comparison analysis showed that for our ensemble and physics goals, the choice of exact distillation is the most advantageous~\citep{Lachini2022}. Also, the smearing radius for this choice is in the order of ${R \approx 1~\mathrm{fm}}$, which is a typical QCD scale.

\section{Spectrum and Finite-volume Analysis}
\label{sec:analysis}

We outline our procedure for calculating scattering parameters here. Using the GEVP data obtained in Sec.~\ref{sec:gevp}, we carefully estimate the systematic effects associated with determining lattice energies. For clarity, we divide the process into two substeps:
\begin{itemize}
    \item[(A)] \emph{Spectrum determination}: finite-volume energy levels for all flavor and irrep quantum numbers $\mathcal{Q}$ are extracted via fits to each GEVP eigenvalue $\lambda^{\mathcal Q}(t)$
     \item[(B)] \emph{Phase shift determination}: resonance parameters are computed via fits of a given phase-shift model to the lattice energy levels
\end{itemize}
Probability distributions are assigned to each of these substeps based on information criteria as specified below. These distributions will be crucial for estimating the systematic uncertainties, which are derived from the spread of the final distribution of the resonance parameters.

In the following sections, we detail how we explicitly apply this strategy to the data generated in our lattice calculation. In Sec.~\ref{sec:spectrumdetermination}, we describe our fits to the lattice data to extract the energy levels and their distributions. In Sec.~\ref{sec:phaseshiftdetermination}, we give details on the determination of the phase shift from the energy spectrum and the models used, and in Sec.~\ref{sec:phaseshiftdetermination} we present how to compute pole parameters from that. In \ref{sec:modelaveraging}, we describe our strategy to assess the systematic error from fit range choices and describe a sampling procedure to estimate the final distribution of phase-shift parameters and pole parameters. We also allow for some variation of the phase shift models in Sec.~\ref{sec:phaseshiftdetermination}.

\subsection{Spectrum Determination}
\label{sec:spectrumdetermination}

For each correlation matrix $C^{\mathcal Q}(t)$ we obtain eigenvalues $\lambda^{(i)}(t)$ by solving the GEVP in Eq.~\eqref{gevpequation}. We remind the reader of the compound index $(i)=\{\mathcal Q, n \}$. We have tested the impact of the choice for the parameter $t_0$ on the RHS of Eq.~\eqref{gevpequation} for all $\mathcal{Q}$ and found that for $t_0 \ge 3a$ the eigenvalues and effective masses are consistent within the increasing statistical noise for higher values of $t_0$. This parameter is set to $t_0=3a$ for the remainder of this work. 

To each eigenvalue $\lambda^{(i)}(t)$, we perform fits of a single-exponential model with parameters $\bm \zeta \equiv \{ Z_{\mathrm{fit}} , E_{\mathrm{fit}}\}$, \ie
\begin{equation}
    \lambda_\mathrm{fit}(t; \bm \zeta) = Z_\mathrm{fit} e^{- t E_\mathrm{fit}}  \,,
    \label{singleexpmodel}
\end{equation}
to each GEVP eigenvalue on \textit{all possible} time ranges $[\tmin^{\mathrm{f}_{(i)}}, \tmax^{\mathrm{f}_{(i)}}]$ fully contained within a \textit{scan range} $[\tstart^{(i)}, \tstop^{(i)}]$, \ie
\begin{equation}
    [\tmin^{\mathrm{f}_{(i)}}, \tmax^{\mathrm{f}_{(i)}}] \subseteq [\tstart^{(i)}, \tstop^{(i)}]  \,,
    \label{fitrange}
\end{equation}
where ${\mathrm{f}_{(i)} \in \{ 1,\, \ldots, \, n^{(i)}_\mathrm{fits}\}}$ labels the fit ranges and $n^{(i)}_\mathrm{fits}$ is the total number of possible fit ranges. We note that, unlike the case of a pseudoscalar two-point function, such as for the pion or kaon, the backward propagating wave has a more complex description, particularly for moving systems~\citep{Asmussen:2022nuk}. This effect has been studied, for example, in the context of $I=2$ $\pi \pi$ scattering~\citep{Dudek:2012gj}. In our work, however, we limit our analysis to the single-exponential case defined earlier, which, as explained in the next sections, describes our data sufficiently well. Given a GEVP eigenvalue and a choice of fit range, the fit corresponds to a minimization of the correlated chi-square
\begin{align}
\begin{split}
    \chi^2_i (\bm \zeta^{\mathrm{f}_{(i)}}) \equiv & \sum_{t,t' \in [\tmin^{\mathrm{f}_{(i)}},\tmax^{\mathrm{f}_{(i)}}]}  \left[ \lambda^{(i)}(t) - \lambda_{\mathrm{fit}}(t; \bm \zeta^{\mathrm{f}_{(i)}}) \right] \times \\
    & \quad \quad  \left( \CovEig^{-1} \right)_{tt'} \left[ \lambda^{(i)}(t') - \lambda_\mathrm{fit}(t'; \bm \zeta^{\mathrm{f}_{(i)}}) \right]
    \label{correlatorchi2}
\end{split}
\end{align}
over the parameters $\bm \zeta$. We propagate statistical errors using the bootstrap method with ${N_\mathrm{boot}=2000}$ samples, and so the $\chi^2$ minimization is repeated at each data bootstrap sample ${b=1, \ldots, N_\mathrm{boot}}$. We perform it an extra time on the gauge average of the eigenvalues, which we label $b=0$. The covariance matrix ${\left(\CovEig\right)_{tt'} \equiv \mathrm{Cov} \left(\lambda^{(i)}(t), \lambda^{(i)}(t') \right)}$ is estimated via the bootstrap method
\begin{align}
    \mathrm{Cov}(u,v) = \frac{1}{N_\mathrm{boot}-1} \sum_{b=1}^{N_\mathrm{boot}} (u_b - \bar u) (v_b - \bar v)  \,,
    \label{generalcovarianceformula}
\end{align}
where the bootstrap estimators are ${\bar u = N_\mathrm{boot}^{-1} \sum_{b=1}^{N_\mathrm{boot}} u_b}$, for a given set of bootstrap samples of the observables $u$ and $v$. This yields the optimal fit parameters $\bm{\zeta}^{\mathrm{f}_{(i)}*}_b$ for eigenvalue $\lambda^{(i)}$ and fit range $\mathrm{f}_{(i)}$. Details on the numerical implementation of this procedure can be found in Appendix~\ref{apx:numericalimplementations}. We only consider fits to at least ${\deltatmin + a}$ consecutive time slices, \ie
\begin{equation}
    \label{defdeltatmin}
    \tmax^{\mathrm{f}_{(i)}} - \tmin^{\mathrm{f}_{(i)}} \ge \deltatmin \, 
\end{equation}
for all $\mathrm{f}_{(i)}$. For later use, the energy $E^{\mathrm{f}_{(i)}}_\mathrm{fit}$ can then be boosted to the c.m. frame using the continuum dispersion relation Eq.~\eqref{continuumdispersionrelation}, in the form 
\begin{equation}
    E^{\mathrm{f}_{(i)}}_{\cm, b} \equiv \sqrt{ (E_{\mathrm{fit}, b}^{\mathrm{f}_{(i)}})^2 - \mathbf P^2 }  \, .
    \label{cmenergyresultdefinition}
\end{equation}

As already anticipated, we restrict the data we analyze in two ways: for each level $(i)$, we determine a range of time slices $[\tstart^{(i)}, \tstop^{(i)}]$ within which we perform fits to all fit ranges allowed by $\deltatmin$. We use a minimal signal-to-noise ratio ($\SNRmin$) condition to determine $\tstop^{(i)}$. More explicitly, for each energy level $(i)$,  we determine $\tstop^{(i)}$ as the earliest time slice for which 
 \begin{equation}
    \label{snr}
    \frac{ \lambda^{(i)}(\tstop^{(i)}) }{ \sigma_{{(i)}}(\tstop^{(i)})} < \SNRmin  \,,
\end{equation}
where $\sigma_{(i)}(t)$ is the statistical uncertainty of the eigenvalue $\lambda^{(i)}(t)$, computed via bootstrap, and $\SNRmin$ is some positive number. All time slices $t>\tstop^{(i)}$ are discarded for the subsequent analysis. We also discard time slices with $t < \tstart^{(i)} = t_0 + a = 4a $ for every level $(i)$. Our procedure then has two remaining parameters, fixed for all ${(i)}$, that can be tuned, namely $\SNRmin$ and  $\deltatmin$. We find that, for the fit ranges given by taking $\SNRmin \ge 5$ and $\deltatmin \ge 5a$, we can fit Eq.~\ref{singleexpmodel} to all the eigenvalues in a well-defined numerical procedure (see Appendix~\ref{apx:numericalimplementations}). Each choice of such hyperparameters (``$\runs$'') defines the dataset in a realization of the analysis, and in Sec.~\ref{sec:results} they are varied to yield our final results.

\begin{figure}[!t]
    \centering
    \includegraphics[width=8.6cm]{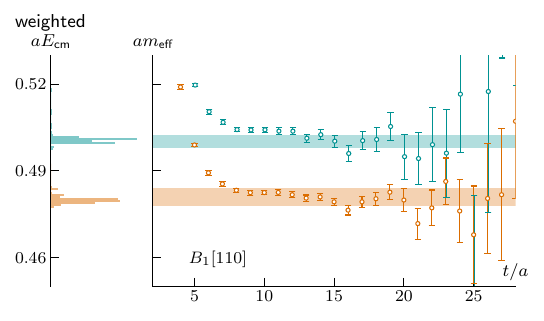}
    \caption{Effective mass of the GEVP eigenvalues considered in the $B_1[110]$ irreps of the $K\pi$ channel. The $w_\corr^{(i)}$-weighted histograms of the central energy fit results to a single exponential are displayed with corresponding colors. Similar plots containing all levels used in this work are shown in Appendix~\ref{apx:gevpeffspectra}. The bands contain the $w_\corr^{(i)}$-weighted statistical error and the fit-range systematic resulting from the model average procedure later described in Sec.~\ref{sec:modelaveraging}, added in quadrature.}
    \label{fig:effmass-histo}
\end{figure}

We further assign an Akaike information criterion ($\AIC$)~\citep{Akaike1974, Akaike1978, Borsanyi2021} to each fit result~\footnote{We also verified that the definition of the $\AIC$ where ${n^\mathrm{data} \to 2 n^\mathrm{data}}$~\citep{Neil2023} yields equivalent results for the purposes of this work.}
\begin{equation}
    \AICc^{(i)} (\mathrm{f}_{(i)}) = \chi^2_i(\bm{\zeta}^{*\mathrm{f}_{(i)}}_{b=0}) + 2 n^\mathrm{par} - n_{\mathrm{f}_{(i)}}^\mathrm{data}  \,,
    \label{eachcorrelatoraic}
\end{equation}
where the number of fit parameters is ${n^\mathrm{par} = 2}$ in the case of the single-exponential form in Eq.~\eqref{gevpequation-eigenvalue} and ${n_{\mathrm{f}_{(i)}}^\mathrm{data} = \tmax^{\mathrm{f}_{(i)}}/a - \tmin^{\mathrm{f}_{(i)}}}/a + 1$ is the number of time slices entering a given fit. In this notation, the number of degrees of freedom is $n^\mathrm{dof}_{\mathrm{f}_{(i)}} = n_{\mathrm{f}_{(i)}}^\mathrm{data} - n^\mathrm{par}$.

For visualization purposes, we compute the effective masses
\begin{equation}
    \label{gevplogmass}
    m^{(i)}_\mathrm{eff}(t) = \log \frac{\lambda^{(i)}(t)}{\lambda^{(i)}(t+1)}
\end{equation}
and their statistical uncertainties via the bootstrap procedure. This is displayed in Fig.~\ref{fig:effmass-histo} for a representative irrep, where we also show the corresponding histograms of central energy results ($b=0$) boosted to the c.m. frame over all possible fit ranges and weighted by
\begin{equation}
    \label{eachcorrelatorweight}
    w_{\corr}^{(i)} \left( \mathrm{f}_{(i)} \right) \propto \exp \left[ -\frac{1}{2} \AICc^{(i)} \left( \mathrm{f}_{(i)} \right) \right]  \, .
\end{equation}
The normalization is chosen such that $w_{\corr}^{(i)}$ can be interpreted as probability distributions, \ie ${\sum_{\mathrm{f}_{(i)}=1}^{n^{(i)}_\mathrm{fits}} w_{\corr}^{(i)} (\mathrm{f}_{(i)}) = 1}$. The $\AIC$ weighting allows to promote fits with a good fit quality and a large number of degrees of freedom. The $w^{(i)}_\corr$-weighted histograms graphically represent the distributions above and their spread provides a systematic uncertainty caused by the fit-range choice at the level of the finite-volume energies (see Eqs.~\eqref{modelavgcentral}-\eqref{modelavgsyst} and subsequent discussion). Such distributions, rather than individual energy levels, are the quantities that will be used in the next sections to propagate a fit-range systematic down to the resonance parameters in Secs.~\ref{sec:modelaveraging},\ref{sec:poleposition}.

\subsection{Phase Shift Determination}
\label{sec:phaseshiftdetermination}

We extracted all fit-range variations of $n^\mathrm{lev}$ finite-volume energy levels from all irreps considered in their respective symmetry channels. Here, $n^\mathrm{lev}=13$ for $K\pi$ scattering and $n^\mathrm{lev}=21$ for $\pi\pi$ scattering. In this section, we refer to a \textit{collection} of finite-volume energies (or fit ranges) as the combination of one representative fit result for each of the $n^\mathrm{lev}$ GEVP eigenvalues, boosted to the c.m. frame via the continuum dispersion Eq.~\eqref{cmenergyresultdefinition}, and bundled together as
\begin{align}
    \{ E^{\mathrm{f}_{(1)}}_\cm,\, E^{\mathrm{f}_{(2)}}_\cm,\, \ldots,\, E^{\mathrm{f}_{(n^\mathrm{lev})}}_\cm \} \equiv \{ E_\cm \}^\mathrm{f},
    \label{FVset}
\end{align}
where we left the bootstrap ($b>0$) and gauge average ($b=0$) index implicit. In Eq.~\eqref{FVset}, the superscript in $E^{\mathrm{f}_{(i)}}_\cm$ indicates that it was the result of a fit with time range $\mathrm{f}_{(i)}$, and $\mathrm{f}$ is a shorthand for a collection of fit ranges ${ \mathrm{f}_{(1)},\, \mathrm{f}_{(2)},\, \ldots ,\, \mathrm{f}_{(n^\mathrm{lev})} }$.

As explained in Sec.~\ref{sec:phaseshift}, the finite-volume energies can be related to the phase shift via the L\"uscher formalism. One can use Eq.~\eqref{pseudophaseqcform} to directly compute the phase shift, and thus the scattering amplitude, at the discrete lattice energies. We are however interested in finding the complex-energy poles of the amplitude. An established procedure is to find suitable energy-dependent functions to model the amplitude and then analytically continue them into the complex plane~\citep{Wilson:2014cna,Rendon:2020rtw}. Given a phase-shift model $\delta^\mdl_1$ with parameters grouped in a vector $\bm \alpha^\mdl$, we can invert the quantization condition in Eq.~\eqref{pseudophaseqcform} to obtain a set of finite-volume energies $\{ E^{\mdl, (i)}_\cm(\bm \alpha^\mdl) \}$.~\footnote{For quantization conditions of type Eq.~\eqref{pseudophaseqcform}, there is a unique $E^{\mdl, (i)}_\cm(\bm \alpha^\mdl)$ for a given $\bm \alpha^\mdl$ and $(i)$. In practice, this involves a numerical root-finding procedure which we solve using the Brent-Dekker algorithm~\citep{Brent1973,Dekker1969} implemented in the GSL library~\citep{GSL}.} Finally, we constrain the phase shift parameters by minimizing~\citep{Guo2013a, Wilson:2015dqa, Erben:2019nmx} 
\begin{align}
\begin{split}
    \chi^2_\PS(\bm{\alpha}^{\mdl, \mathrm{f}})& = \sum_{i,j=1}^{n^\mathrm{lev}} \left[ E^{\mathrm{f}_{(i)}}_\cm - E^{\mdl, (i)}_\cm(\bm{\alpha}^{\mdl, \mathrm{f}}) \right] \\
    & \times (\CovPS^{-1})_{ij}^\mathrm{f} \left[ E^{\mathrm{f}_{(j)}}_\cm - E^{\mdl, (j)}_\cm(\bm{\alpha}^{\mdl, \mathrm{f}}) \right]
    \label{spectrumchisquared}
\end{split}
\end{align}
on \textit{every} bootstrap sample and on the gauge average, for a given collection $\{ E_\cm \}^\mathrm{f}$. The covariance matrix ${\left( \CovPS \right)^\mathrm{f}_{ij} \equiv \mathrm{Cov} \left( E^{\mathrm{f}_{(i)}}_\cm ,  E^{\mathrm{f}_{(j)}}_\cm \right)}$ is computed via Eq.~\eqref{generalcovarianceformula}. Numerical details can be found in Appendix~\ref{apx:numericalimplementations}.

In summary, for a given phase-shift model $\delta^\mdl_1$ and a given $\{ E_\cm \}^\mathrm{f}$ obtained from some combination of fit ranges $\mathrm{f}$, we find the optimal $\bm{\alpha}^{\mdl, \mathrm{f}*}$ which minimizes \eqref{spectrumchisquared}. In line with Eq.~\eqref{eachcorrelatoraic}, we assign~\footnote{For simplicity of notation, we drop the $*$ from the optimal fit parameters minimizing a chi-square.}
\begin{equation}
    \label{phaseaic}
    \AICps^{{\mdl}}\left(\mathrm{f}\right) = \chi_\PS^2(\bm{\alpha}^{\mdl, \mathrm{f}}_{b=0}) + 2 n_\mdl^\mathrm{par} - n^\mathrm{lev}
\end{equation}
to such a fit, where $n^\mathrm{par}_\mdl$ is the number of phase-shift parameters and $n^\mathrm{lev}$ is the number of finite-volume levels, the latter fixed in our analysis. We further define the phase-shift weight
\begin{equation}
    \label{phaseweight}
    w_\PS^\mdl \left( \mathrm{f} \right) \propto \exp \left[ -\frac{1}{2} \AICps^\mdl \left(\mathrm{f}\right) \right]
\end{equation}
in similarity to Eq.~\eqref{eachcorrelatorweight}.  

In this work, we employ two different phase-shift models and study the systematic effect that this choice has on the resonance description. In the following, we introduce the particular models employed.

\subsubsection{Breit-Wigner $( \BW )$}

Given Eq.~\eqref{tmatrixdef}, the Breit-Wigner model parametrizes the scattering amplitude as~\citep{BreitWigner1936, PDG2024}
\begin{equation}
    \tmat{1}^\BW \! \left(\sqrt{s}\right) = \frac{\Gamma^\BW \sqrt{s}}{(m)^2 - s - i \Gamma^\BW \sqrt{s}}
    \label{amplitudebw}
\end{equation}
in terms of the Breit-Wigner mass $m$ and width $\Gamma^\BW$. Anticipating an eventual analytical continuation, we use the invariant-mass notation, related to the c.m. momentum via Eq.~\eqref{CMenergies}.

In $P$-wave, the amplitude above can be expressed in terms of a phase shift as
\begin{equation}
    p_\cm^3 \cot \delta^\BW_1(\sqrt{s}) = \frac{6 \pi}{g^2} (m^2 - s) \sqrt{s}\,, \quad \bm \alpha^\BW = [g, m]\,.
    \label{bwmodel}
\end{equation}
The energy-dependent Breit-Wigner width is related to the effective coupling $g$ via
\begin{equation}
    \Gamma^\BW(s) = \frac{g^2}{6\pi} \frac{p_\cm^3}{s}\,,
\end{equation}
which ensures the expected relativistic behavior at the two-particle threshold. The specific form of the width introduces a model dependence that can be relevant for resonances far from the narrow-width approximation~\citep{PDG2024}.

\subsubsection{Effective Range Expansion $( \ERE )$}

The second model we employ to parametrize the phase shift is the \textit{effective range expansion}~\citep{Bethe:1949yr}, which is an expansion of $p_\cm^3 \cot \delta_1$ in powers of $p_\cm^2$, \ie
\begin{equation}
    \label{eremodel}
    p_\cm^3 \cot \delta^\ERE_1(\sqrt{s}) = \frac{1}{a_1} + \frac{r_1}{2} p_\cm^2 \,, \quad \bm \alpha^\ERE = [a_1, r_1]\,,
\end{equation}
to first order. It features the so-called scattering length $a_1$ and the effective range $r_1$, which in $P$-wave have dimensions of volume and inverse length, respectively. This model can describe resonant, as well as weakly attractive or repulsive scattering~\citep{Wilson:2014cna}, and contains the appropriate angular momentum behavior expected for $P$-wave.

\subsection{Pole Parameters}
\label{sec:poleposition}

We also compute the resonance pole positions by substituting the optimal phase shift parametrizations found via Eq.~\eqref{spectrumchisquared} back into the elastic scattering amplitude from Eq.~\eqref{tmatrixdef},
\begin{equation}
    \tmat{1}^\mdl(\sqrt{s}) = \frac{1} {\cot \delta^\mdl_1(\sqrt{s}) \rvert_{\bm \alpha^\mdl} - i } \,,
\end{equation}
and analytically continue it into the complex-$\sqrt{s}$ plane. Now $\tmat{1}^\mdl$ is explicitly defined on the complex plane and will feature the expected multi-sheet structure encoded by Eq.~\ref{CMenergies}. The identification of the poles on the second Riemann (unphysical) sheet provides a model-independent definition of resonances.

We compute the so-called resonance \textit{pole parameters}
\begin{equation}
    \sqrt{s_\mathrm{pole}} = M - \frac i2 \, \Gamma \,,
    \label{poleposdef}
\end{equation}
located on the unphysical sheet, where ${\mathrm{Im}~p_\cm < 0}$. This defines the \textit{pole mass} $M$ and \textit{pole width} $\Gamma$ of a resonance. In the Breit-Wigner case, we implement the pole-finding by minimizing the quantity~\footnote{We use the \textit{Neldermead} algorithm~\citep{NELDERMEAD,NLopt} for this step.}
\begin{equation}
    \mathcal{F}(p_\cm) \equiv \Big| \, \tmat{1} \left( \sqrt{s} (p_\cm) \right) \, \Big|^{-2}
\end{equation} 
in the complex-$p_\cm$ plane, restricted to the unphysical sheet. In lattice units, the minimum $p^\mathrm{pole}_\cm$ is found to a precision ${a \epsilon_\mathrm{pole}= 10^{-7}}$ and then confirmed to be a zero of $\tmat{1}^{-1}$ by checking that $\mathrm{Re}~\tmat{1}^{-1}$ and $\mathrm{Im}~\tmat{1}^{-1}$ change sign for at least one pair of vertices around the solution. \ie for at least one pair in
\begin{equation}
    \Big( \, \mathrm{Re}~p^\mathrm{pole}_\cm \pm \epsilon_\mathrm{pole} \,, \, \mathrm{Im}~p^\mathrm{pole}_\cm \pm \epsilon_\mathrm{pole} \, \Big) \, .
\end{equation}
On the other hand, the first-order effective range model can be written as an order-three polynomial equation which can be exactly solved. We then search for the unphysical-sheet pole closest to the physical scattering line and take that as the resonance pole position from the $\ERE$ model. 

In the fashion of Sec.~\ref{sec:phaseshiftdetermination}, we determine the pole positions $(M)^{\mdl , \mathrm{f}}$ and $(\Gamma)^{\mdl , \mathrm{f}}$ not only for each phase-shift model $\mdl$, but also for each underlying fit range combination $\mathrm{f}$. These will be combined into a single result for the pole parameters as follows.

\subsection{Model Averaging Procedure}
\label{sec:modelaveraging}

We now give our prescription for determining the final statistical and systematical uncertainties of the quantities introduced in the previous sections. In particular, we estimate the systematical error on such quantities due to the underlying choice of fit range in the GEVP eigenvalues fits. This type of estimation through a data-driven procedure was already performed in other lattice QCD calculations~\citep{BMW:2014pzb,Borsanyi2021}, and here we formulate it to the workflow typical of the Lüscher method. This can also be viewed as a model-selection problem, not only at the level of the phase shift but in the form of correlator fit range selection~\citep{Jay2021, Neil2023}.

We note first that, given a choice of fit ranges for all levels and a phase-shift model, an ideal fit procedure can be imagined such that it yields the phase-shift parameters through a \textit{single fit}. This would be achieved by a global correlated minimization of a suitable $\chi^2_\mathrm{ideal}$, leading to some $\AIC_\mathrm{ideal}$. The spread of the $e^{-\AIC_\mathrm{ideal}/2}$-weighted distribution of phase-shift parameters caused by the variation of the fit range choices could then be taken as a systematic error. However, the associated global covariance matrix cannot be reliably inverted with the available statistics, and thus realizing such fit is not a feasible task.

As an approximation to this ideal fit, our procedure instead blocks the fitting strategy into two substeps (\ref{sec:spectrumdetermination} and \ref{sec:phaseshiftdetermination}), which are iterated to produce distributions of the finite-volume spectrum and phase-shift parameters, respectively. The $\AIC_\mathrm{ideal}$ envisioned in the paragraph above is then approximated by the \textit{total $\AIC$}
\begin{align}
    \AICt^{\mdl} \left(\mathrm{f}\right) \equiv \AICps^{\mdl}\left(\mathrm{f}\right) + \sum_{i=1}^{n^\mathrm{lev}} \AICc\left(\mathrm{f}\right)
    \label{totalAIC}
\end{align}
for a given phase-shift model $\delta_1^\mdl$, where $\mathrm{f}$ labels the underlying collection of correlator fit ranges. Note that Eq.~\eqref{totalAIC} assumes that the correlations between time slices within the same GEVP eigenvalue ($\CovEig$) and between the energies ($\CovPS$) are the dominant ones, while the remaining are neglected. This criterion defines the weights
\begin{equation}
    w^{\mdl}_\mathrm{t} \left(\mathrm{f}\right) \propto \exp{ \left[ -\frac 12 \AICt^{\mdl}\left(\mathrm{f}\right) \right] } \,,
    \label{totalaicweight}
\end{equation}
which implies
\begin{equation}
    w^{\mdl}_\mathrm{t} \left(\mathrm{f}\right) = w_\corr (\mathrm{f}) \, w^{\mdl}_\PS (\mathrm{f})
    \label{wtwpswcorr}
\end{equation}
with
\begin{equation}
    w_\corr (\mathrm{f}) \equiv \prod^{n_\mathrm{lev}}_{i=1} w_\corr^{(i)} (\mathrm{f}_{(i)})  \,.
    \label{totalcorrweight}
\end{equation}
Again, the normalization is chosen such that $w^{\mdl}_\mathrm{t}$ is a well-defined probability. The weight $w^{\mdl}_\mathrm{t}$ assigns a probabilistic interpretation to the results, \eg allowing the calculation of expectation values
\begin{equation}
    \langle \alpha^\mdl \rangle_\mathrm{t} = \sum_\mathrm{f} \alpha^{\mdl, \mathrm{f}} w_\mathrm{t}^\mdl (\mathrm{f}) \,,
    \label{fullexpectationvalue}
\end{equation}
where the summation symbol without a superscript represents the sum over \textit{all} fit-range combinations. The subscript ``$\mathrm{t}$'' of the expectation value denotes that it is taken with respect to the total weight $w_\mathrm{t}^{\mdl}$.

Analyzing the spread of $w^{\mdl}_\mathrm{t}$ due to fit range choices is challenging, as it lacks an analytical expression. On the other hand, a brute-force method would involve performing phase-shift fits for all possible fit-range combinations, which is impractical given the vast number of combinations, $n^{(1)}_\mathrm{fits} \times n^{(2)}_\mathrm{fits} \times \ldots \times n^{(n^\mathrm{lev})}_\mathrm{fits} \sim {\mathcal{O}(10^{20})-\mathcal{O}(10^{40})}$. Instead, the approach we use is \textit{importance sampling}~\citep{NumericalRecipes2007}, where fit-range samples are drawn from a \textit{proposal} distribution and then reweighted to match $w^{\mdl}_\mathrm{t}$.

\begin{figure}[!t]
    \centering
    \includegraphics[width=8.6cm]{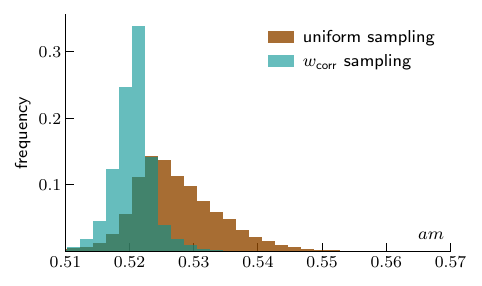}
    \caption{Histograms of the $K\pi$ Breit-Wigner mass (see Eq.~\ref{bwmodel}) obtained from $\Nscan=50{,}000$ fit-range samples drawn uniformly and from the $w_{corr}$ distribution of the underlying correlator fit ranges.
    We emphasize that the phase-shift fit weights do not enter the construction of these histograms. We also stress that the spread of the ``uniform sampling'' histogram is not indicative of the systematic uncertainty on $a m$, since fits are uniformly included irrespective of fit quality. This histogram is included only to illustrate the space of fits sampled. Final histograms are given in Appendix~\ref{apx:individualmodelpsparhisto}, where the reweighting step was performed.}
    \label{fig:uniformvsAICsampling}
\end{figure}

In principle, even a uniform distribution could be used to generate sample proposals. However, importance sampling is most effective when its samples target the region where the true distribution is concentrated. In our approach, $w_\corr$ selects fit ranges that produce energies close to the true lattice values, making it a more appropriate proposal distribution compared to drawing samples with equal probability. This can be understood as a suppression of the large number of phase-shift fits with large $\AICps$ that would have virtually no contribution to the final result. In Fig.~\ref{fig:uniformvsAICsampling}, we illustrate how the distribution of a phase-shift parameter over fit-range samples is more localized when drawn from $w_\corr$ instead of a uniform distribution. In particular, $\mathcal{O}(10^5)$ uniform drawings were necessary before a single reasonable $\chi^2_\PS/n^\mathrm{dof} \approx 1$ was found, compared to $\mathcal{O}(10^2)-\mathcal{O}(10^3)$ for $w_\corr$ sampling.  After reweighting to obtain $w^{\mdl}_\mathrm{t}$, we find that nearly all fits from the uniform distribution are suppressed, whereas those from $w_\corr$ are frequently enhanced.

\begin{figure*}[!th]
    \centering
    \includegraphics[width=17.2cm]{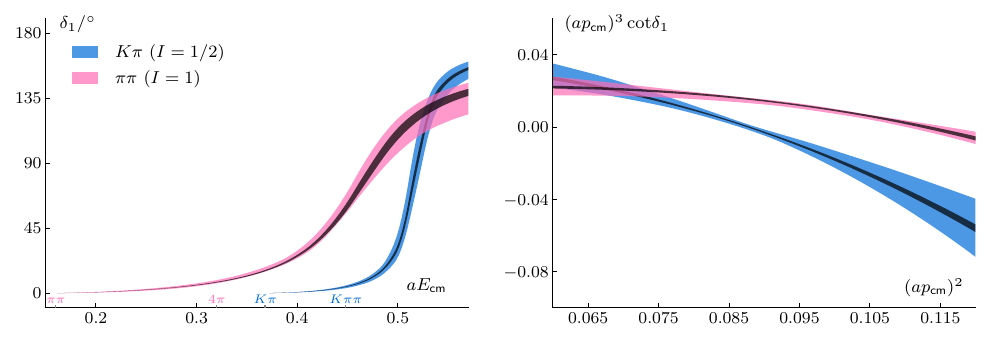} 
    \caption{Model average of the (real part of) the phase shifts (left) and $(a p_\cm)^3\cot \delta_1$ (right), with statistical (black) and data-driven systematical (colorful) bands. Both bands are obtained by the procedure in Sec.~\ref{sec:modelaveraging} applied to such quantities on each energy and momentum-cubed bin, accounting for the $\BW, \ERE$ model variation, as well as for the GEVP fit-range variation and the hyperparameter choices (``$\runs$'') from Table~\ref{tab:hyperparametervariation}.}
    \label{fig:phaseshifts}
\end{figure*}

We proceed with $w_\corr$ as the proposal distribution, which in the limit of an infinite number of samples yields
\begin{align}
    \begin{split}
        & \langle \alpha^\mdl \rangle_\mathrm{t} = \Big\langle \alpha^{\mdl, \mathrm{s}} \, \frac{w^\mdl_\mathrm{t}(\mathrm{s})}{w_\corr(\mathrm{s})} \Big\rangle_\corr \,,
        \label{importancesampling}
    \end{split}
\end{align}
where $\mathrm{s}$ is a fit-range sample and $\langle \rangle_\corr$ denotes the expectation value with respect to $w_\corr$. In practice, we adopt the following procedure to draw a finite number of samples:
\begin{itemize}
    \item for every $(i)$, draw a fit range $\mathrm{f}_{(i)}$ with probability given by $w_{\corr}^{(i)}(\mathrm{f}_{(i)})$~\footnote{See Appendix~\ref{apx:inversesampling} for the explicit implementation.}
    \item build the \textit{collection sample} 
    \begin{equation}
        {\{ \mathrm{f}_{(1)},\, \mathrm{f}_{(2)},\, \ldots ,\, \mathrm{f}_{(n^\mathrm{lev})} \}} \equiv \mathrm{s}
        \label{fitrangesample}
    \end{equation}
    \item repeat $\Nscan$ times to get $\mathrm{s}^1 ,\, \mathrm{s}^2 ,\, \ldots ,\, \mathrm{s}^{\Nscan}$
\end{itemize}
The result of this procedure is $\Nscan$ energy samples distributed according to $w_\corr$. For each fit-range collection sample $\mathrm{s}^k$, we write the corresponding combination of finite-volume energies as
\begin{equation}
    \{ E_\cm \}^{\mathrm{s}^k}, \quad k = 1 ,\, \ldots ,\, \Nscan \,,
    \label{poolofenergies}
\end{equation}
where the index $k$ counts the samples. On each $\{ E_\cm \}^{\mathrm{s}^k}$, the phase-shift parameter for a given model $\delta^\mdl_1$ is computed via minimization of Eq.~\eqref{spectrumchisquared} and labeled as $\alpha^{\mdl,\mathrm{s}^k}$. The pole parameters follow from the phase-shift parameters, \cf Sec.~\ref{sec:poleposition}, and are labeled in an analogous way, \ie $M^{\mdl,\mathrm{s}^k}$ and $\Gamma^{\mdl,\mathrm{s}^k}$.

In the following, we present our prescription for estimating the statistical and systematic uncertainties of the phase-shift parameters from the $w^\mdl_\mathrm{t}$ distribution. Due to the employed importance sampling, the scattering results from the various collections in Eq.~\eqref{poolofenergies} need to be reweighted by $w^\mdl_\mathrm{t}/w_\corr = w^\mdl_\PS$, as indicated in Eq.~\eqref{importancesampling}. 
For $\Nscan$ fit-range collection samples $\mathrm{s}^k$ distributed according to $w_\corr$, the weights $w^\mdl_\PS$ can be computed as in Eq.~\eqref{phaseweight} and used to define $\hat w^\mdl_\PS \equiv w^\mdl_\PS / \sum_{k=1}^{\Nscan} w^\mdl_\PS (\mathrm{s}^k)$.~\footnote{Note that the normalization constant going into Eq.~\eqref{importancesampling} is estimated from importance sampling to be $\Nscan^{-1} \sum_{k=1}^{\Nscan} w^\mdl_\PS (\mathrm{s}^k)$.} 
Given a phase-shift parameter $\alpha^\mdl$, we define the estimators:
\begin{itemize}
    \item Central value :
        \begin{equation}
            \label{modelavgcentral}
            \hat{\alpha}^\mdl_{b=0} \equiv \sum_{k=1}^{\Nscan} \alpha_{b=0}^{\mdl, \mathrm{s}^k} \hat{w}^\mdl_\PS (\mathrm{s}^k)\,,
        \end{equation}
        where the circumflex ($\hat\ $) denotes a $\hat w^\mdl_\PS$-\textit{weighted} mean and $b=0$ indicates the gauge-average result
    \item Statistical uncertainty :
        \begin{equation}
            \label{modelavgstat}
            \sigma_{\alpha^\mdl, \stat} \equiv \mathrm{Cov} \left(\hat{\alpha}^\mdl, \hat{\alpha}^\mdl \right)^{1/2} \,,
        \end{equation}
        \ie the bootstrap standard deviation of the $\hat  w^\mdl_\PS$-\textit{weighted} means, \cf \eqref{generalcovarianceformula}
    \item Systematical uncertainty interval :
        \begin{equation}
            [ \alpha^{\mdl}_{\sys-}, \alpha^{\mdl}_{\sys+} ] \,,
            \label{modelavgsyst}
        \end{equation} 
        such that $\alpha^{\mdl}_{\sys-}$ and $\alpha^{\mdl}_{\sys+}$ are respectively the $2.1\%$ and $97.9\%$ $\hat w^\mdl_\PS$-\textit{weighted} percentiles of $\alpha_{b=0}^{\mdl, \mathrm{s}^k}$
\end{itemize}

For the pole parameters $M^{\mdl,\mathrm{s}^k}$ and $\Gamma^{\mdl,\mathrm{s}^k}$, we define a total-$\AIC$ weight in the same way as in Eq.~\eqref{totalaicweight}, now normalized not only over the $\Nscan$ fit-range samples but also the phase-shift models. We then compute the central value, statistical and systematical uncertainties as above, but with the replacement ${\sum_{k=1}^{\Nscan} \to \sum_\mdl \sum_{k=1}^{\Nscan}}$ and its associated normalization factors.

We empirically observe that the final results are reasonably stable for a number of fit-range samples greater than $\Nscan = 50{,}000$, which we adopt throughout the following analysis. Across the $\Nscan = 50{,}000$ fit-range combinations we have also determined the value of $\chi^2_\PS/n_\mathrm{dof}$ that is closest to $1$ for each of the two fit models and the various sampling realizations described in Sec.~\ref{sec:results}. For $K \pi$, we find $\chi^2_\PS/n_\mathrm{dof} = 0.99-1.00$ with $n_\mathrm{dof} = 11$. Similarly, for $\pi \pi$, we find $\chi^2_\PS/n_\mathrm{dof} = 1.15-1.86$ with $n_\mathrm{dof} = 19$. 

An analogous procedure can be employed for the finite-volume energies, which we treat as a separate result to the scattering determination. In this case, each level comprises only $\mathcal{O}(10^2)$ results, one for each possible fit range, and thus sampling is not necessary. They are given in a similar form as Eqs.~\eqref{modelavgcentral},\eqref{modelavgstat} and \eqref{modelavgsyst}, but with the replacement ${\sum_{k=1}^{\Nscan} \hat w^\mdl_\PS \to \sum_{k=1}^{n^{(i)}_\mathrm{fits}} (\mathrm{s}) w_\corr^{(i)}} (\mathrm{f}^k_{(i)})$, \ie
\begin{equation}
    \hat E^{(i)} = \sum_{k=1}^{n^{(i)}_\mathrm{fits}} w_\corr^{(i)} (\mathrm{f}^k_{(i)}) E^{\mathrm{f}^k_{(i)}}\, ,
    \label{eq:energycorrweighting}
\end{equation}
so that no phase-shift information is used in their determination. 

In any of the quantities above, we define a symmetrized version of the central value as the center of the systematic interval and thus the symmetrized systematic uncertainty as half of the interval length. We use the asymmetric version in figures and the symmetrized one otherwise.

\section{Results and Discussions}
\label{sec:results}

\begin{figure*}[!h]
    \centering
    \includegraphics[width=17.2cm]{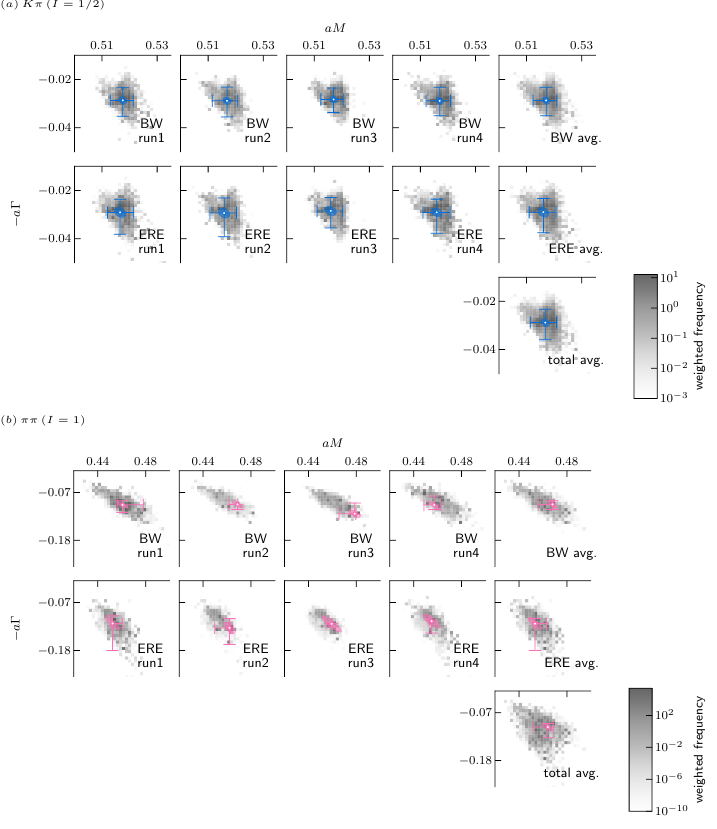}
    \caption{Breakdown of the data-driven systematic spread of the resonance pole-positions in the $K\pi$ (top) and $\pi\pi$ (bottom) amplitudes extracted. Variations are taken over the $\BW, \ERE$ phase-shift models and all the hyperparameter choices in Table~\ref{tab:hyperparametervariation}, labeled as ``$\run{1}$'',``$\run{2}$'',``$\run{3}$'' and ``$\run{4}$''. The systematic interval is denoted by the cross errorbar and the statistical error ellipse shows the correlation between $M$ and $-\Gamma$. The two-dimensional histogram is the $\AICps$-weighted frequency over corresponding finite-volume energy samples and it is plotted on a log scale. The normalization of the weights is taken such that the histograms on all panels are comparable to each other.}
    \label{fig:breakdownpole}
\end{figure*}

As detailed in Sec.~\ref{sec:spectrumdetermination}, the allowed eigenvalue fit ranges determining the finite-volume energies are controlled by the hyperparameters $(\SNRmin,\deltatmin)$. We account for their systematic impact on the pole parameters by repeating the complete analysis workflow for the four choices shown in Table~\ref{tab:hyperparametervariation}. The variation of these runs is a measure of the systematic uncertainty in our final result and it is detailed in Figs.~\ref{fig:breakdownpole} and Table~\ref{tab:individualpolepos}. All four ``runs'' and also the variation over the two phase-shift models can be combined into a single pole position result by performing the procedure from Sec.~\ref{sec:modelaveraging} over the different realizations of $(\SNRmin,\deltatmin)$ and the models $\BW$ and $\ERE$. This effectively leads to the replacement ${\sum_{k=1}^{\Nscan} \to \sum_\runs \sum_\mdl \sum_{k=1}^{\Nscan}}$ in the weighting of the pole parameters when compared to Eq.~\eqref{modelavgcentral}. Our (symmetrized) final results in lattice units for the two scattering channels considered in this work are
\begin{align}
\begin{split}
    &a M_{K^*}     = 0.5160 (10)_\stat (47)_\dd     \\
    &a \Gamma_{K^*}= 0.0296 (11)_\stat (63)_\dd       
    \label{overallkstarlattice}
\end{split}
\end{align}
and
\begin{align}
\begin{split}
    &a M_\rho     = 0.4603 (28)_\stat(87)_\dd        \\
    &a \Gamma_\rho= 0.1112 (58)_\stat(164)_\dd \,,     
    \label{overallrholattice}
\end{split}
\end{align}
where the uncertainty in the first bracket is our statistical uncertainty, indicated by the $\stat$ subscript, and the second bracket corresponds to the \textit{data-driven systematic} uncertainty, indicated by $\dd$, which in addition to the variation over the four ``$\runs$'' and the two phase-shift models also includes the fit-range variation explained in detail in Secs.~\ref{sec:spectrumdetermination},\ref{sec:modelaveraging}. We also introduced the notation $M_\rho, \Gamma_\rho$ for the pole mass and width of the $\rho$ and analogously $M_{K^*}, \Gamma_{K^*}$ for the pole mass and width of the ${K^*}$. Using a similar procedure, we also determine the statistical and data-driven systematic bands for both the (real part of the) phase-shift and $p_\cm^3 \cot \delta_1$, shown in Fig.~\ref{fig:phaseshifts}. The corresponding individual results for the phase-shift parameters are listed in Table~\ref{tab:phaseshiftpars}.

\begin{table}[!t]
    \centering
    \setlength{\tabcolsep}{0.5em}{\renewcommand{\arraystretch}{1.5}
    {   
    \begin{tabular}{|c|c|c|}
    \hline
    $\run{}$                 & ${\SNRmin}$   & ${\deltatmin/a}$    \\ \hline
    $\mathrm{1}$             & $8$           & $5$                \\ \hline
    $\mathrm{2}$             & $5$           & $7$                \\ \hline
    $\mathrm{3}$             & $7$           & $7$                \\ \hline
    $\mathrm{4}$             & $6$           & $6$                \\ \hline
    \end{tabular}
    }}
    \caption{Variation of the hyperparameters $(\SNRmin, \deltatmin)$ used in the resonance determination (\cf Eqs.~\eqref{defdeltatmin},\eqref{snr}).}
    \label{tab:hyperparametervariation}
\end{table}

Note that the statistical fluctuations due to the sampling method are probed by the hyperparameter variations of Table~\ref{tab:hyperparametervariation}, which are taken into account by our data-driven uncertainties. This is demonstrated in Fig.~\ref{fig:breakdownpole} by comparing the different ``runs'' to the averaged result. We further check that those fluctuations are under control by repeating the entire calculation with different bootstrap and fit-range sampling random seeds.

The finite-volume energies are also model-averaged across the various realizations of $(\SNRmin,\deltatmin)$, similarly implying the replacement ${\sum_{k=1}^{n^{(i)}_\mathrm{fits}} \to \sum_\runs \sum_{k=1}^{n^{(i)}_\mathrm{fits}}}$ in Eq.~\eqref{eq:energycorrweighting}. The result from this is exemplified in Figs.~\ref{fig:effmass-histo}, and fully shown in Fig.~\ref{fig:nonint_modelavg_spectrum} and Appendix~\ref{apx:gevpeffspectra}. For convenience, we also provide a comparison to other ways of obtaining the spectrum in Appendix~\ref{apx:otherspectrum}.

We do not perform a dedicated scale setting in this work, but instead use the value of the inverse lattice spacing, $a^{-1}=\valueainverse$, computed in Ref.~\citep{RBC:2014ntl}. We assume this measurement to be statistically uncorrelated from our resonance determination and use it for quoting results in physical units. We indicate the statistical error coming from $a^{-1}$ in the results given in physical units as a separate contribution labeled as ``$\scale$''.

In the following, we estimate the sources of systematic uncertainties that are not addressed directly in this work and quote them in the physical unit results. They are summarized together with the data-driven systematics in Table~\ref{tab:errorbudget}.

\textit{Discretization}: We work with only one lattice spacing, meaning that we cannot study the continuum limit. We instead assign a discretization systematic for expected $\mathcal{O}(a^2)$ effects from naive power counting~\citep{Boyle:2022lsi}. Given a conservative estimate of $\Lambda_{QCD} \approx 400$ MeV, we get $(a \Lambda_{QCD})^2 = 5\%$, which dominates all the other following systematics.

\textit{Dispersion relation}: The fits of pseudoscalar two-point correlators to the dispersion relation with an adjustable slope yield deviations from the continuum relation at the percent level (see Sec.~\ref{sec:dispersion}). We assume that when boosted to the c.m. frame using the continuum relation in Sec.~\ref{sec:phaseshift}, the interacting lattice energies used in the quantization condition will carry such an effect. We comment that this effect should in principle be covered by the discretization uncertainty estimated above and is sub-leading enough that it has no effect on the error budget of the final result whether we include it or not.

\clearpage

\onecolumngrid
\begin{table*}[!t]
    \centering
    \setlength{\tabcolsep}{0.5em}{\renewcommand{\arraystretch}{1.5} 
    {
        \begin{tabular}{c|c|c}
            \hline
            \textbf{Source}                 & $\bm M$             & $\bm \Gamma$                                \\\hline\hline
            scale setting                   & $0.2\%$               & $0.2\%$                                       \\\hline
            data-driven                     & $(0.9\%)_{K\pi}, (1.9\%)_{\pi\pi}$ & $(21.3\%)_{K\pi}, (14.8\%)_{\pi\pi}$ \\\hline
            discretization                  & $5\%$                 & $5\%  $                                       \\
            dispersion relation             & $1\%$                 & $1\%$                                         \\
            quark mass                      & $1\%$                 & $1\%  $                                       \\
            inelastic thresholds            & $1\%$                 & $1\%$                                         \\
            partial-wave truncation         & $1\%$                 & $1\%$                                         \\
            residual finite-volume effects  & $0.1\%$               & $0.1\%$                                       \\\hline
        \end{tabular}
        }}
    \caption{Error budget used to quote final results on the resonance pole parameters. We control the data-driven systematic, while all other entries below it are estimates. For our final result, we quote the statistical ($\stat$), data-driven ($\dd$) and scale setting ($\scale$) uncertainties separately. We combine all other systematics ($\oth$) into one combined error of $6\%$.}
    \label{tab:errorbudget}
\end{table*}
\twocolumngrid

\textit{Quark Mass}: The tuning of bare quark masses in the domain-wall action and the exact isospin symmetry inflict a percent level difference between the pion and kaon masses in our lattice compared to their physical values~\citep{RBC:2014ntl}. Results from chiral perturbation theory~\citep{Nebreda2010} suggest the highest deviation on pole parameters is also at the percent level on the $K^*$ and $\rho$ widths. For near-physical pions and kaons, the variation in the resonance masses is even milder. As mentioned in Sec.~\ref{sec:ensemble}, another uncertainty in our calculation stems from using the zM\"obius approximation to the M\"obius DWF action in the valence sector. We thus estimate the total effect on the pole parameters to be at the percent level.

\textit{Residual finite-volume effects}: The presence of a periodic finite box leads to exponentially suppressed corrections due to virtual pion effects, which are discarded in the Lüscher method~\citep{Luscher:1985dn,Luscher:1986pf}. At our $m_{\pi} L \approx 3.8$, there is a percent level error on the asymptotic expansion leading to the quantization condition~\citep{Kim:2005gf}. More directly, the associated deviation on pole parameters was previously reported to be at the permille level in effective theory studies~\citep{Albaladejo2013,Chen2012}.

\textit{Partial-wave truncation}: The quantization condition Eq.~\eqref{pseudophaseqcform} assumes that $\delta_{\ell > 1} = 0$~\citep{Kim:2005gf}, which is correct up to leading corrections coming from $F$-waves ($\ell=3$). Due to the generic partial-wave suppression~${\sim (p_\cm)^{-4}}$ in relation to $\ell=1$, we take this as a percent effect. This is backed up by previous studies on $\rho$ and $K^*$ which observed only small contributions from higher partial waves subducing into lattice irreps~\citep{Dudek:2012xn,Andersen:2018mau,Wilson:2014cna}.

\textit{Inelastic thresholds}: From unitarity, the scattering amplitude accumulates contributions to its imaginary part above the $K\pi\pi$ and $\pi\pi\pi\pi$ thresholds. The corresponding phase shifts thus become complex-valued in those regions. We argue that ignoring such thresholds inflicts an error on the $\rho$ and $K^*$ parameters that can be taken as a systematic uncertainty. In a physical point calculation, the strongest evidence comes from the experimental branching fractions of resonances~\citep{PDG2024}
\begin{align}
\begin{split}
    & \Gamma_{K^* \to K\gamma}  \approx 3 \times 10^{-3} \\
    & \Gamma_{K^* \to K\pi\pi} < 7 \times 10^{-4}        \\
    & \Gamma_{K^* \to K\pi\pi\pi} \  \text{(unknown)}
    \label{branchingratioskstar}
\end{split}
\end{align}
and
\begin{align}
\begin{split}
    & \Gamma_{\rho \to \pi\gamma}       \approx 5 \times 10^{-3} \\
    & \Gamma_{\rho \to \pi\eta}         \approx 6 \times 10^{-3} \\
    & \Gamma_{\rho \to \pi\pi\gamma}    \approx 10^{-2}          \\
    & \Gamma_{\rho \to \pi\pi\pi\pi}    < 2 \times 10^{-3} \,.
    \label{branchingratiosrho}
\end{split}
\end{align}
Through the optical theorem, the imaginary part of the phase shift due to a certain threshold will be of the order of the branching fraction on the corresponding decay channel. From the infinite-volume side, we can argue that the contribution on $\Gamma$ due to decays into $K\pi\pi$ and $\pi\pi\pi\pi$ is $\lesssim 10^{-3}$ at the corresponding threshold energies. This view is reinforced by the theoretically expected chiral and phase-space suppressions of the $K\pi \to K\pi\pi$ and $\pi\pi \to \pi\pi\pi\pi$ amplitudes in comparison to their $2 \to 2$ counterparts. Furthermore, the $K\pi \to K\pi\pi$ process corresponds to an anomalous term of the Wess-Zumino form in the chiral Lagrangian~\citep{Wess1971,Meissner1988}. Besides chirally suppressed with $\sim F^{-5}_{\pi} \approx (90\mev)^{-5}$, such a term depends on very specific external momenta configurations due to a contraction with the Levi-Civita symbol, leading to an additional kinematic suppression. In fact, due to the larger decay fraction to final states with a photon, one could argue that the inclusion of electromagnetic effects should be addressed even before multi-pion effects are detectable.

We are using the two-particle quantization condition, which does not take into account finite-volume effects from diagrams where three or more propagators can simultaneously go on-shell in the $s$-channel~\citep{Luscher:1986pf,Kim:2005gf}. The result is that this method ignores the imaginary part attained by the phase shift above $K\pi\pi$ and $\pi\pi\pi\pi$ thresholds, though we expect it to be small at physical quark masses.

On the other hand, due to the small coupling to the $K^*$ and $\rho$ resonances, \cf \eqref{branchingratioskstar} and \eqref{branchingratiosrho}, the $K\pi\pi$ and $\pi\pi\pi\pi$ energies in their respective channels will lie very close to their non-interacting energies, except when near level crossings. In our main analysis, we do not use the states with level-crossings near ${L = 48}$ (see Fig.~\ref{fig:nonint_modelavg_spectrum}). In principle, the GEVP matrices including $K\pi\pi$ and $\pi\pi\pi\pi$ correlators can be computed to directly estimate the size of the off-diagonal terms, but this is out of the scope of this work. Nevertheless, we expect the generic volume scaling $\braket{0 | O_{K\pi} | K\pi\pi} \sim L^{-3}$ and $\braket{0 | O_{\pi\pi} | \pi\pi\pi\pi} \sim L^{-6}$, which suggests that three and four-particle operators would have a small effect on the two-particle levels we do include in the final analysis. A more detailed understanding of the effect of crossing the three-particle threshold could be achieved by applying the extended finite-volume formalism of Ref.~\citep{\TwoToThree}. See also Refs.~\citep{\ThreeParticles} for more general reviews on progress for describing three particles in a finite volume.

A summary of our error budget can be found in table~\ref{tab:errorbudget}. Our largest uncontrolled uncertainty is due to the discretization effects and it dominates our extra error budget. We combine all the systematic effects discussed above into an overall $6\%$ uncertainty on our results quoted in physical units. We denote this uncertainty as ``$\oth$''.

With all these uncertainties taken into account, our final result for the pole parameters in physical units is
\begin{widetext}
\begin{align}
\begin{split}
    M_{K^*}       &= 893 (2)_\stat (8)_\dd  (54)_\oth (2)_\scale \mev \\
    \Gamma_{K^*}  &= 51  (2)_\stat (11)_\dd (3)_\oth  (0)_\scale \mev   
\end{split}
\end{align}
and
\begin{align}
\begin{split}
    M_\rho       &= 796 (5)_\stat  (15)_\dd  (48)_\oth (2)_\scale \mev \\
    \Gamma_\rho  &= 192 (10)_\stat (28)_\dd (12)_\oth (0)_\scale \mev\,. 
\end{split}
\end{align}
We can combine the three systematic errors $()_\dd ()_\oth ()_\scale$ by quadrature into one $()_\sys$ error, yielding
\begin{align}
\begin{split}
    M_{K^*}       &= 893 (2)_\stat (54)_\sys \mev \\
    \Gamma_{K^*}  &= 51  (2)_\stat (11)_\sys \mev   
\end{split}
\end{align}
and
\begin{align}
\begin{split}
    M_\rho       &= 796 (5)_\stat  (50)_\sys \mev \\   
    \Gamma_\rho  &= 192 (10)_\stat (31)_\sys \mev   \,.
\end{split}
\end{align}
\end{widetext}

\bigskip
\section{Conclusions}
\label{sec:conclusions}

In this work, we computed resonance parameters corresponding to the physical $K^*(892)$ and $\rho(770)$ particles using first-principles lattice QCD simulations with a physical pion mass. We estimated the associated fit range systematics by developing a data-driven technique applied to Lüscher-type calculations. In the data-driven systematic we also accounted for the phase-shift model dependency, using an effective-range parametrization and a Breit-Wigner one. We compare the results from our work, with all uncertainties taken into account, to the experimental value provided by PDG in Table \ref{tab:finalResult}.

\begin{table}[!b]
    \centering
    \setlength{\tabcolsep}{0.5em}{\renewcommand{\arraystretch}{1.5}{
    \begin{tabular}{c|c|c}
        \hline
        & this work $[\mev]$ & PDG $[\mev]$ \\
        \hline
        $M_{K^*}$           & $893 (2)_\stat (54)_\sys $     & $890(2)$     \\
        $\Gamma_{K^*}/2$    & $26  (1)_\stat (6)_\sys   $    & $25.6(1.2)$       \\
        \hline
        $M_\rho$            & $796 (5)_\stat (50)_\sys $     & $761 - 765$   \\
        $\Gamma_\rho/2$     & $96  (5)_\stat (15)_\sys  $    & $71 - 74$     \\
    \end{tabular}
    }}
    \caption{Comparison of the experimental values of the pole-position parameters provided by PDG~\citep{Pelaez:2020uiw, Colangelo:2001df, Garcia-Martin:2011nna,PDG2024} to the ones from this work, with all statistical $()_\stat$ and systematic $()_\sys$ uncertainties taken into account. The experimental values for the $\rho$ channel are given as a range rather than a central value with an uncertainty. Note that by convention the resonance width $\Gamma$ is twice the imaginary part, but for ease of comparison to experimental data we provide half the width or the imaginary part of the pole position.}
    \label{tab:finalResult}
\end{table}

Within the given uncertainties, our resonance-pole-position results are in agreement with the experimental values. Our theoretical uncertainties are completely dominated by systematic effects rather than statistical ones. When considering the separate sources of systematic errors, then the pole width $\Gamma$ obtains the largest systematic effects from the data-driven $()_\dd$ analysis, whereas other effects are sub-dominant. The pole masses $M$ on the other hand have a similarly-sized data-driven uncertainty in the $K^*$ channel and a much smaller data-driven uncertainty in the $\rho$ channel when compared to $\Gamma$, despite the mass being much larger than the width in both channels. The dominating systematic effect in the pole masses therefore stems from the fact that our computation only uses a single lattice spacing. We account for this via an estimate of a $5\%$ error on the central value. By extending this work to more lattice spacings we would be able to considerably lower this conservatively chosen uncertainty, and obtain uncertainties on the masses with a similar size to the experimental uncertainties.

Our $\rho$ resonance parameters are also in good agreement with other lattice QCD calculations \citep{ExtendedTwistedMass:2019omo,Fischer:2020yvw} which give a physical-point result via a chiral-continuum extrapolation. There are no such lattice QCD calculations yet of the $K^*$ resonance at physical kinematics or with a continuum limit being taken.

Moving forward, it would be desirable to repeat the calculation at additional lattice spacings which would help with the systematic uncertainties for the pole masses. Reducing the relatively large uncertainty in the resonance widths stemming from the data-driven analysis poses a more difficult question, as that is considerably sensitive to the regions of the phase-shift where the interactions are weaker and thus where the finite-volume effects are smaller. 

A crucial way forward is not only in the better control of systematic uncertainties of our results but also in the application of similar finite-volume formalisms for more complex quantities involving weak-transition matrix elements. These include the transition amplitudes of $\pi \gamma \to \pi \pi$ and $K \gamma \to K \pi$, both previously studied at larger-than-physical pion masses, as well as semileptonic decays of heavy mesons, like $B \to \rho \ell \nu$ and $B_{(s)} \to K^* \ell^+ \ell^-$, which are currently being investigated in the lattice QCD community~\citep{Leskovec:2022ubd, Leskovec:2024pzb}.

\bigskip
The data entering this publication was generated using the openly available lattice QCD software packages Grid~\citep{Boyle2015} and Hadrons~\citep{Hadrons2023}. The supporting data for this article are openly available from the CERN document server \citep{boyle_2024_vy9x7-bzn92}.

\begin{acknowledgments}
The authors thank the members of the RBC and UKQCD Collaborations for the helpful discussions and suggestions. N.P.L., F.E. and A.P. kindly thank Mike Peardon for the invaluable discussions. N.P.L. additionally thanks André Baião Raposo for the discussions. This work used the DiRAC Extreme Scaling service (Tursa / Tesseract) at the University of Edinburgh, managed by the Edinburgh Parallel Computing Centre on behalf of the STFC DiRAC HPC Facility (www.dirac.ac.uk). The DiRAC service at Edinburgh was funded by BEIS, UKRI and STFC capital funding and STFC operations grants. DiRAC is part of the UKRI Digital Research Infrastructure. P.B. has been supported in part by the U.S. Department of Energy, Office of Science, Office of Nuclear Physics under the Contract No. DE-SC-0012704 (BNL). P.B. has also received support from the Royal Society Wolfson Research Merit award WM/60035. M.M. gratefully acknowledges support from STFC in the form of a fully funded PhD studentship. A.P. \& F.E. received funding from the European Research Council (ERC) under the European Union's Horizon 2020 research and innovation programme under grant agreement No 757646. N.P.L. \& A.P. received funding from the European Research Council (ERC) under the European Union's Horizon 2020 research and innovation programme under grant agreement No 813942. N.P.L. also acknowledges support from the U.K. Science and Technology Facilities Council (STFC) [grant numbers ST/T000694/1, ST/X000664/1]. F.E. has received funding from the European Union’s Horizon Europe research and innovation programme under the Marie Sk\l{}odowska-Curie grant agreement No. 101106913. M.T.H. and F.J. are supported by UKRI Future Leader Fellowship MR/T019956/1. A.P., M.T.H., V.G. \& F.E. were supported in part by UK STFC grant ST/P000630/1, and A.P., M.T.H. \& V.G. additionally by UK STFC grants ST/T000600/1 \& ST/X000494/1. A.P. is additionally partially supported by a long-term Invitational Fellowship from the Japan Society for the Promotion of Science.
\end{acknowledgments}

\appendix

\onecolumngrid

\section{Effective GEVP Spectra}
\label{apx:gevpeffspectra}

\begin{figure*}[!h]
    \centering
    \includegraphics[width=17.2cm]{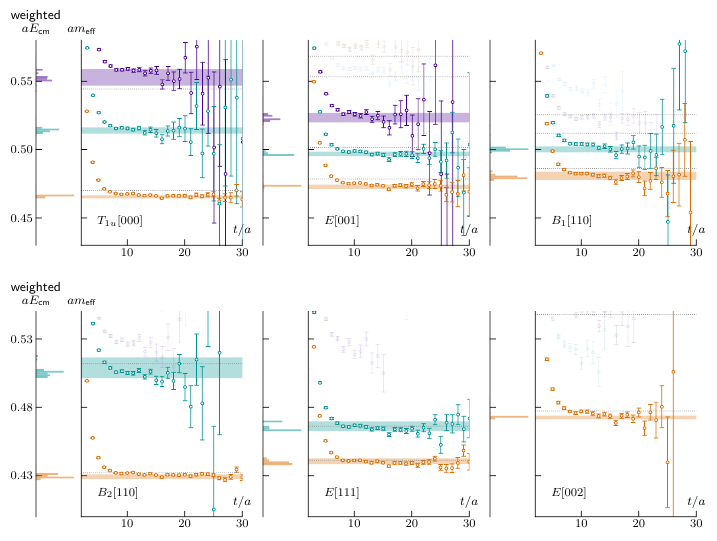}
    \caption{Effective masses, $w_\corr^{(i)}$-weighted histograms and model-averaged c.m. finite-volume energy plots for all the considered irreps in the $K\pi$ $(I=1/2)$ analysis. The bands contain statistical, fit-range and hyperparameter variations. The faint colorful data represent levels that were not used in the scattering analysis and the faint gray dashed lines are the two-particle $K\pi$ non-interacting c.m. energies with units of momentum corresponding to the ones used in the two-bilinear operators, \cf Table~\ref{tab:irrepmomenta}.}
    \label{fig:effmass_spectrum_kpi}
\end{figure*}

\begin{figure*}[!h]
    \centering
    \includegraphics[width=17.2cm]{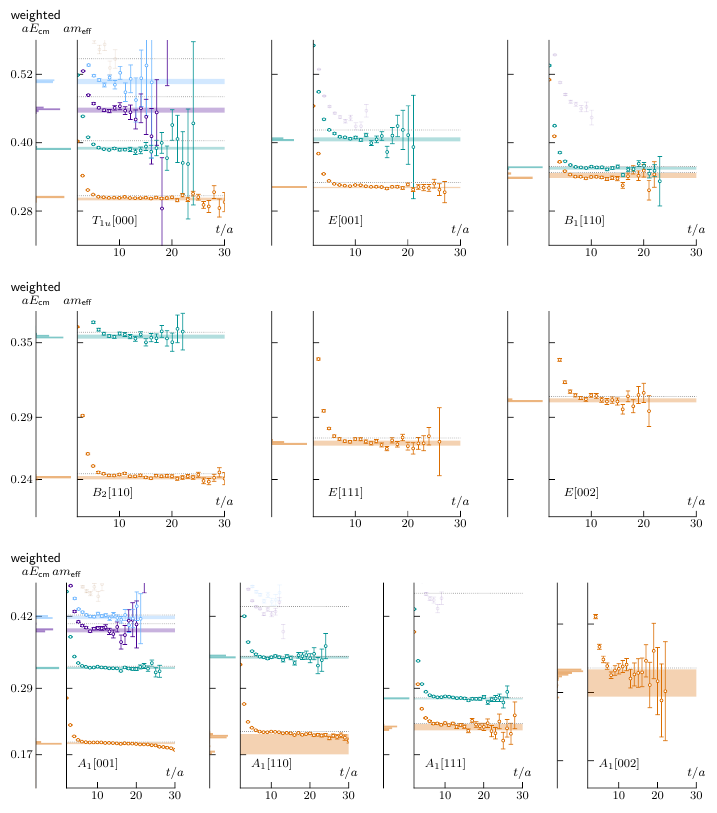}
    \caption{Effective masses, $w_\corr^{(i)}$-weighted histograms and model-averaged c.m. finite-volume energy plots for all the considered irreps in the $\pi\pi$ $(I=1)$ analysis. The bands contain statistical, fit-range and hyperparameter variations. The faint colorful data represent levels that were not used in the scattering analysis and the faint gray dashed lines are the two-particle $\pi\pi$ non-interacting c.m. energies with units of momentum corresponding to the ones used in the two-bilinear operators, \cf Table~\ref{tab:irrepmomenta}.}
    \label{fig:effmass_spectrum_pipi}
\end{figure*}

We summarize in Figs.~\ref{fig:effmass_spectrum_kpi},\ref{fig:effmass_spectrum_pipi} the effective masses Eq.~\eqref{gevplogmass} computed from the GEVP eigenvalues $\lambda^{(i)}$ on each lattice irrep (see Sec.~\ref{sec:gevp}), for illustration purposes. The faint dashed lines are energies of non-interacting $K\pi$ and $\pi\pi$ systems, each carrying the momenta of the two-bilinear operators used in the GEVP, boosted to the c.m. frame, \cf Eq.~\eqref{nonintCMformula} and Table~\ref{tab:irrepmomenta}. Through the horizontal histograms, we depict the frequency of the c.m.-boosted energies within the $n^{(i)}_\mathrm{fits}$ single-exponential fit results, $E^{(i)}_\cm$, weighted by $w_\corr^{(i)}$, according to the procedure in Sec.~\ref{sec:spectrumdetermination}. The bands are the model-averaged result according to the prescription given in Sec.~\ref{sec:modelaveraging}, over the hyperparameter variations (``$\runs$'') of Table~\ref{tab:hyperparametervariation} (see also Eq.~\eqref{eq:energycorrweighting}). Such bands contain both the statistical error and the systematic due to fit-range and hyperparameter variations, all summed in quadrature. Note that these uncertainties intervals do not go directly into the Lüscher analysis, instead the sampling procedure of Sec.~\ref{sec:modelaveraging} is employed.

\onecolumngrid

\clearpage

\section{Alternative Spectrum Weighting}
\label{apx:otherspectrum}

In Appendix~\ref{apx:gevpeffspectra} above, we show our preferred result for the finite-volume spectrum as determined directly from the GEVP, \ie via the use of the $w_\corr^{(i)}$-weighting in Eq.~\eqref{eq:energycorrweighting}.

\begin{figure*}[b!]
    \includegraphics[width=10cm]{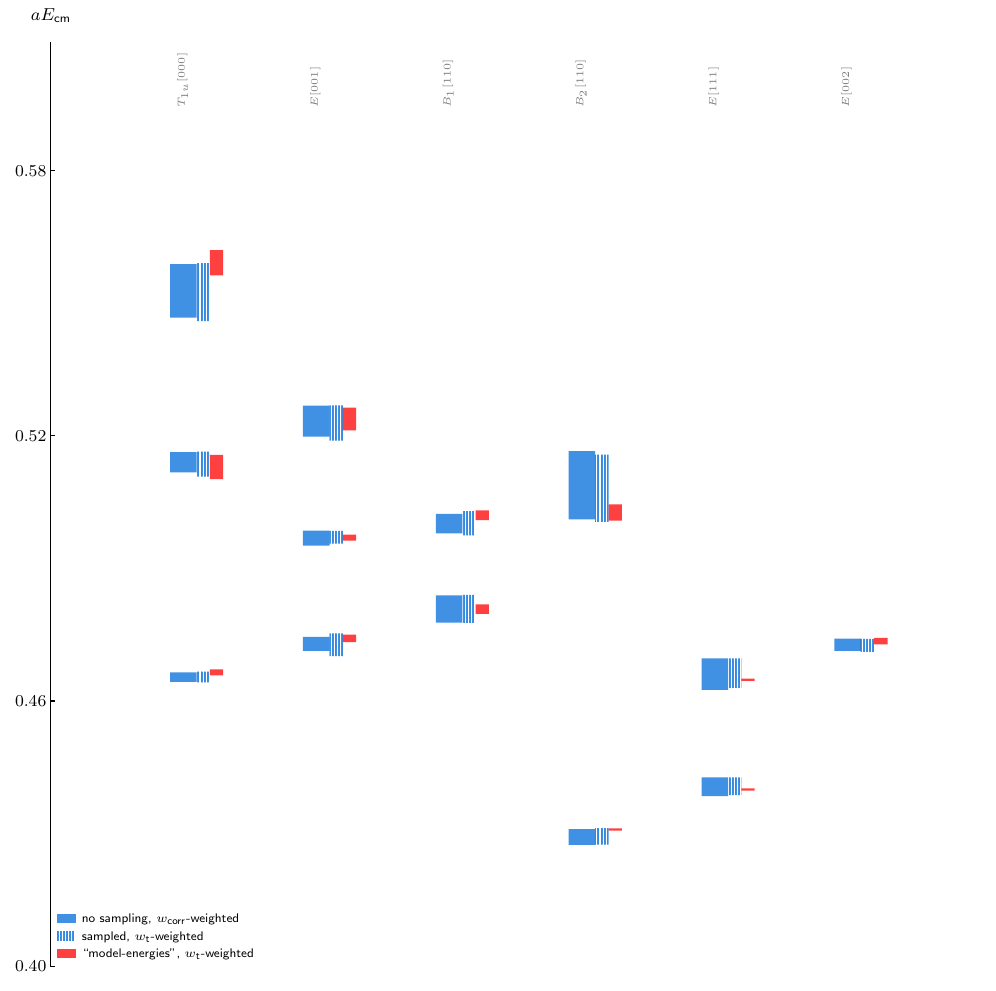}
    \caption{$K\pi$ finite-volume spectrum as determined from: GEVP eigenvalues only as in Eq.~\eqref{eq:energycorrweighting} (``$\hat E^{(i)}$'', left full rectangles); GEVP eigenvalues and $w_\mathrm{t}$ weights as in Eq.~\eqref{eq:spec-wt}(``$\hat E^{(i)}_\mathrm{t}$'', middle hatched rectangles); model results as in Eq.~\eqref{eq:spec-inverse} (``$\hat{\mathcal{E}}^{(i)}$'', right solid rectangles). The statistical and data-driven systematics were computed in analogy to Eqs.~\eqref{modelavgstat},\eqref{modelavgsyst} and added in quadrature.} 
    \label{fig:speccomparison_kpi}
\end{figure*}

\begin{figure*}[b!]
    \centering
    \includegraphics[width=10cm]{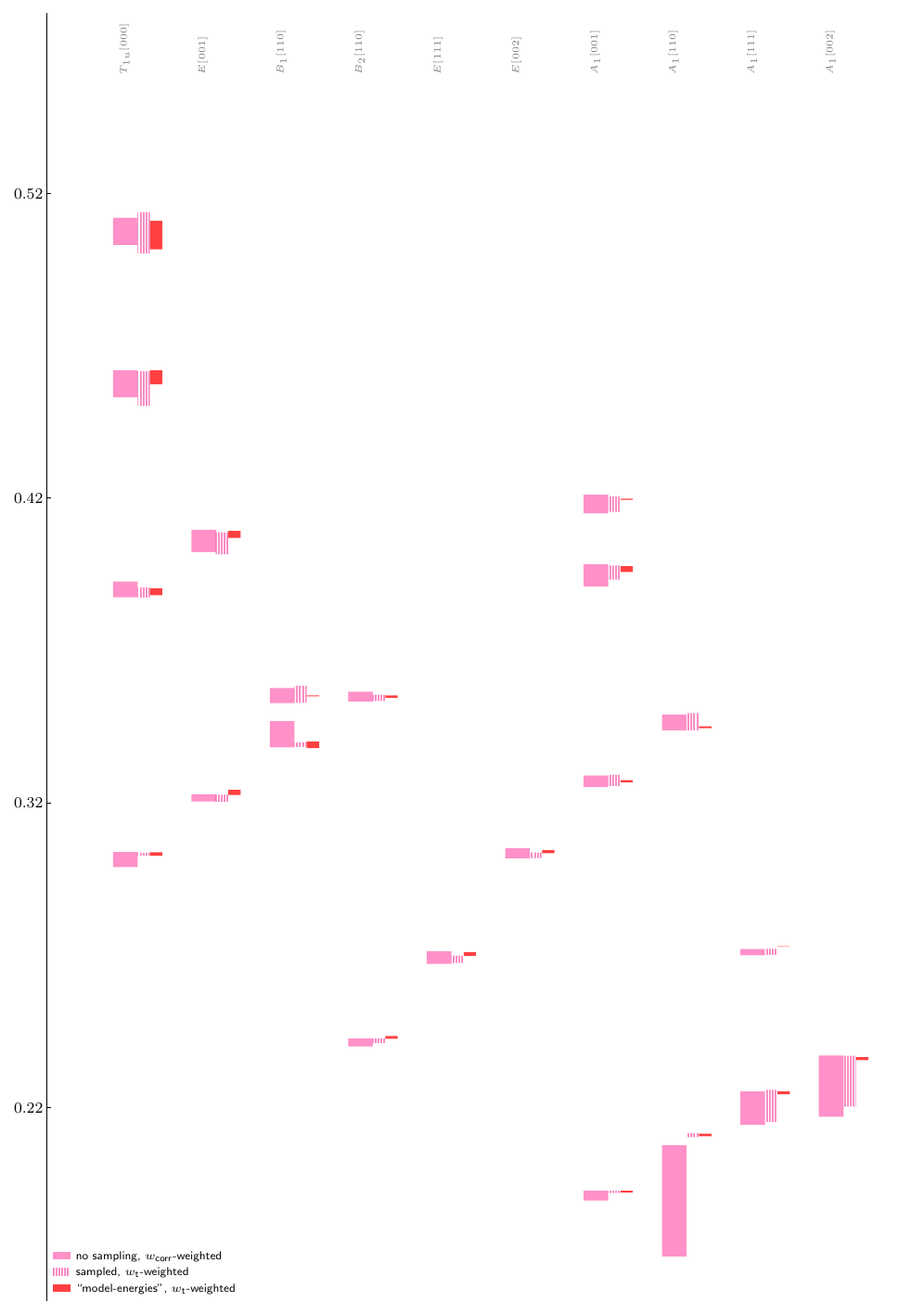}
    \caption{The same as in Fig.~\ref{fig:speccomparison_kpi}, but for the $\pi\pi$ spectrum.}
    \label{fig:speccomparison_pipi}
\end{figure*}

An alternative method is to consider the weighting by $w_\mathrm{t}$, which incorporates to each energy the phase-shift fit quality on each fit-range sample. This can be written over all fit-range combinations as $\langle E^{(j)} \rangle_\mathrm{t} = \sum_\mathrm{f} E^{(j)} \left( \mathrm{f}_{(j)} \right) w_\mathrm{t}(\mathrm{f})$, and estimated via importance sampling by
 \begin{equation}
    \hat E^{(i)}_\mathrm{t} \equiv \sum_{\mdl} \sum_{k=1}^{\Nscan} \hat w_\PS (\mathrm{s}^k, \mdl) E^{\mathrm{s}^k_{(i)}}\, ,  
    \label{eq:spec-wt}
\end{equation}
which is depicted in Figs.~\ref{fig:speccomparison_kpi},\ref{fig:speccomparison_pipi} as the middle hatched rectangles.

Still another perspective on this is to compute the so-called ``interacting energies'' (or ``fit-model energies''), which we denote as $\hat{\mathcal{E}}^{(i)}$. These correspond to $E^{\mdl, (i)}_\cm(\bm{\alpha}^{\mdl})$ in Eq.~\eqref{spectrumchisquared}, evaluated at the $\chi^2_\PS$ minimum. Their full model-averaged value is given by
\begin{equation}
    \hat{\mathcal E}^{(i)} \equiv \sum_{\mdl} \sum_{k=1}^{\Nscan} \hat w_\PS (\mathrm{s}^k, \mdl) \  \mathcal E^{(i)} (\bm{\alpha}^{\mdl, \mathrm{s}^k})\, .
    \label{eq:spec-inverse}
\end{equation}
We left the average over ``$\runs$'' implicit, and uncertainties can be computed similarly to Eqs.~\eqref{modelavgstat},\eqref{modelavgsyst}.

\vfill
\vfill

\clearpage

\twocolumngrid
\section{Wick contractions}
\label{apx:wick}

Here, we show the explicit Wick contractions for each correlation function used in the GEVP before applying any lattice-irrep projection (see Sec.~\ref{sec:operatorprojection}). The interpolators used are listed in Eqs.~\eqref{rhokstarinterpolators}-\eqref{twobilinearmomprojection}, with the addition of a normalization factor $1/\sqrt{2}$ into $O_{\pi\pi}, O_{K\pi}$ and $O_{\pi^0}$.

We use a distillation-based diagrammatic notation inspired in Ref.~\citep{Morningstar:2011ka}. Each $\mathcal{M}^{(f)}_\Gamma(\bf p, t)$ represents a meson field~\eqref{mesonfieldrhophi} of quark flavour $f \in \{ l, s\} $, Dirac matrix $\Gamma$ and momentum $\bf p$. A line connecting two meson fields indicates a matrix product (in dilution indices) between those fields in the reverse order of the arrow so that the lines always originate on a $\varrho$ and end on a $\varphi$. A closed loop represents a trace over the dilution indices.
\bigskip

\begin{figure}[h]
    \centering
    \includegraphics[width=0.5\linewidth]{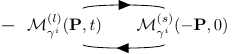}
    \caption{$K\pi$ $(I=1/2)$ -- Diagram in $\braket{O_{K^{*+}}(\mathbf P,t) \, O_{K^{*+}}(\mathbf P,0)^{\dagger}}_F\,$.}
    \label{kstarkstardiagram}
\end{figure}
\bigskip

\begin{figure}[h!]
    \centering
    \includegraphics[width=0.6\linewidth]{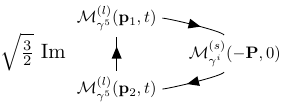}
    \caption{$K\pi$ $(I=1/2)$ -- Diagram in $\braket{O_{K \pi}(\mathbf p_1,\mathbf p_2,t) \, O_{K^{*+}}(\mathbf P,0)^{\dagger}}_F\,$. }
    \label{kpikstardiagram}
\end{figure}
\bigskip

\begin{figure}[h!]
    \centering
    \includegraphics[width=\linewidth]{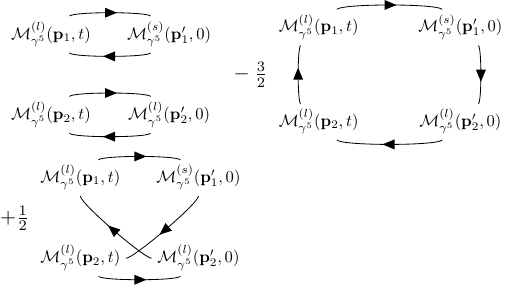}
    \caption{$K\pi$ $(I=1/2)$ -- Diagrams and respective factors going into $\braket{O_{K \pi}(\mathbf p_1,\mathbf p_2,t) \, O_{K \pi}(\mathbf p_1',\mathbf p_2',0)^{\dagger}}_F\,$.}
    \label{kpikpidiagram}
\end{figure}
\bigskip

\begin{figure}[h!]
    \centering
    \includegraphics[width=0.5\linewidth]{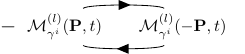}
    \caption{$\pi\pi$ $(I=1)$ -- Diagram in $\braket{O_{\rho^+}(\mathbf P,t) \, O_{\rho^+}(\mathbf P,0)^{\dagger}}_F\,$.}
    \label{rhorhodiagram}
\end{figure}
\bigskip

\begin{figure}[h!]
    \centering
    \includegraphics[width=1\linewidth]{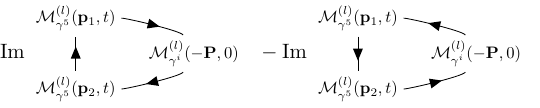}
    \caption{$\pi\pi$ $(I=1)$ -- Diagrams and respective factors going into $\braket{O_{\pi \pi}(\mathbf p_1,\mathbf p_2,t) \, O_{\rho^+}(\mathbf P,0)^{\dagger}}_F\,$.}
    \label{pipirhodiagram}
\end{figure}
\bigskip

\begin{figure}[ht!]
    \centering
    \includegraphics[width=1\linewidth]{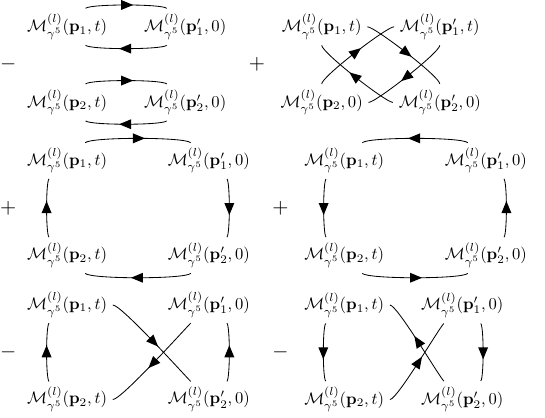}
    \caption{$\pi\pi$ $(I=1)$ -- Diagrams and respective factors going into $\braket{O_{\pi \pi}(\mathbf p_1,\mathbf p_2,t) \, O_{\pi \pi}(\mathbf p_1',\mathbf p_2',0)^{\dagger}}_F\,$.}
    \label{pipipipidiagram}
\end{figure}
\bigskip

We compute the diagrams above for all possible source times and average the ones with equal source-sink separation, represented here by translating the source to the temporal origin. We impose that the correlator matrix should be exactly symmetric and only compute one side of the off-diagonal elements. The diagrams in Figs.~\ref{kpikstardiagram},\ref{pipirhodiagram} are purely imaginary, and we choose an arbitrary global phase that yields them real.

\section{Phase-shift parameter histograms}
\label{apx:individualmodelpsparhisto}

\begin{figure}[h!]
    \centering
    \includegraphics[width=0.9\linewidth]{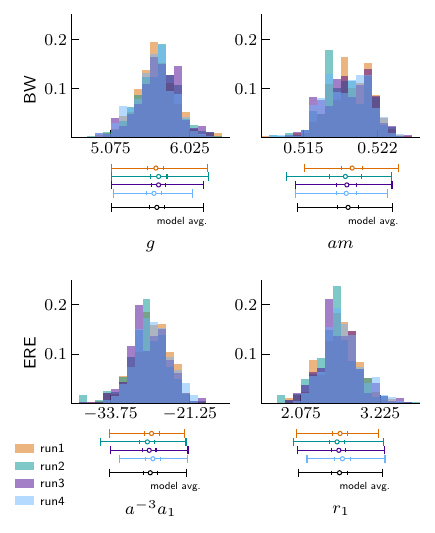} 
    \caption{Phase-shift parameter histograms for ${K\pi \ (I=1/2)}$ over the models in Sec.~\ref{sec:phaseshiftdetermination} and the variations (``$\runs$'') of hyperparameters in Table~\ref{tab:hyperparametervariation}.}
    \label{fig:ps_histo_kpi}
\end{figure}

\begin{figure}[h!]
    \includegraphics[width=0.9\linewidth]{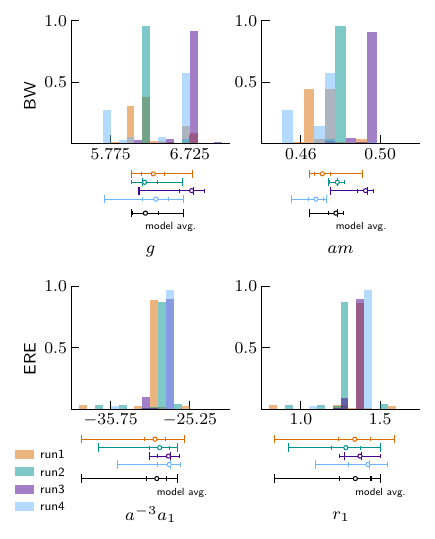}
    \caption{Phase-shift parameter histograms for ${\pi\pi \ (I=1)}$ over the models in Sec.~\ref{sec:phaseshiftdetermination} and the variations (``$\runs$'') of hyperparameters in Table~\ref{tab:hyperparametervariation}.} 
    \label{fig:ps_histo_pipi}
\end{figure}

In Figs.~\ref{fig:ps_histo_kpi},\ref{fig:ps_histo_pipi} we depict the distribution of the phase-shift parameters as histograms stemming from eigenvalue fit-range combinations sampled via $w_\corr$ and weighted by $\hat w_\PS$ from the respective phase-shift model (see Sec.~\ref{sec:modelaveraging}). The results given by error bars contain statistical and fit-range systematic uncertainties as described in Sec.~\ref{sec:modelaveraging}, and they match in color the histogram containing the respective $96\%$ systematic spread. These values correspond to a non-symmetrized version of the results going into Table~\ref{tab:phaseshiftpars}. The model-averaged value is indicated at the bottom in black.

\section{Numerical implementations}
\label{apx:numericalimplementations}

\subsection{Chi-square minimization}
\label{apx:chisquareminimization}

The numerical chi-square minimization of Eq.~\eqref{correlatorchi2} is carried out on a multi-step procedure, where we first perform a preconditioning at a sloppy precision which is then refined in a final step. In the former, we use the global algorithm \textit{GN\_CRS2}~\citep{CRS2, NLopt} followed by \textit{Minuit}~\citep{MinuitZenodo,MinuitAcademic} to minimize \eqref{correlatorchi2} \textit{only} at the gauge-average of the eigenvalues~($b=0$), with a precision of $\epsilon_{\mathrm{precon.}} = 10^{-3}$  and ignoring the off-diagonal components of the covariance matrix, $\CovEig$, for each level labeled by $(i)$. This produces a single initial guess which is then fed into \textit{Minuit} to perform the minimization on \textit{every} bootstrap sample with a precision $\epsilon_{\mathrm{final}} = 10^{-5}$, now using the full covariance $\CovEig$. This is implemented within \textit{LatAnalyze}, an open-source library for lattice data analysis~\citep{Latan2020}.

We perform the numerical minimization of Eq.~\eqref{spectrumchisquared} in a similar two-step procedure as with the eigenvalue fits. However, in order to guarantee fit stability, we perform the very last minimization using the local \textit{SLSQP}~\citep{SLSQP, NLopt} algorithm, just after \textit{Minuit}. In both procedures, we require all the intermediate steps to be completely warning-free.

\subsection{Inverse transform sampling}
\label{apx:inversesampling}

We wish to draw a fit range $\mathrm{f}_k$ from the pool
\begin{equation}
    \{ \mathrm{f}_k \} \,, \quad k = 1, 2, \ldots , n^{(i)}_\mathrm{fits} \,, 
\end{equation}
with probability given by the discrete distribution $w_{\corr}^{(i)} ({\mathrm{f}_k})$. In practice, this is done by first computing the discrete cumulative distribution function
\begin{equation}
    \mathrm{cdf}_{\corr}^{(i)} (\mathrm{f}_k) = \sum_{k' = 1}^{k} w_{\corr}^{(i)} (\mathrm{f}_{k'}) \,,
\end{equation}
thus satisfying $\mathrm{cdf}_{\corr}^{(i)} ( \mathrm{f}_{n^{(i)}_\mathrm{fits}} )= 1$. Then, by drawing a uniformly distributed pseudo-random number $u \in [0,1]$, the smallest $k$ such that $u \leq \mathrm{cdf}_{\corr}^{(i)}(\mathrm{f}_k)$ is drawn with the desired probability.

In the procedures of Sec.~\ref{sec:modelaveraging}, we use different pseudo-random number generator seeds for each of the energy levels but set them to be equal over different realizations of the analysis, \ie different choices of hyperparameters in Table~\ref{tab:hyperparametervariation}.

\section{Results for individual models}
\label{apx:individualmodelnumbers}

In this appendix, we list the numerical values for each individual model fit result described in Sec.~\ref{sec:phaseshiftdetermination} and graphically shown in Figs.~\ref{fig:breakdownpole},\ref{fig:ps_histo_kpi} and \ref{fig:ps_histo_pipi}. The statistical error and fit-range systematics are shown in that order. The result is symmetrized by taking the central value as the center of the systematic interval presented in Sec.~\ref{sec:modelaveraging}. The model-averaged results (``model avg.'') are taken over the hyperparameter choices in Table~\ref{tab:hyperparametervariation}.

\begin{table}[!h]
\centering
    \setlength{\tabcolsep}{0.5em}{\renewcommand{\arraystretch}{1.5}{
    \centering
    \begin{tabular}{ccc|c|c}
    \cline{4-5}
    & &     & $K\pi$ $(I=1/2)$         & $\pi\pi$ $(I=1)$  \\ \hline\hline
    \multicolumn{1}{c|}{\multirow{8}{*}{\rotatebox{90}{$\BW$}}}  & \multirow{2}{*}{\rotatebox{90}{$\run{1}$}} & \multicolumn{1}{|c|}{$g$ } & $5.66(10)(58)$     & $6.39(14)(37)$    \\
    \multicolumn{1}{c|}{}                                      &                                           & \multicolumn{1}{|c|}{$am$} & $0.5195(10)(44)$   & $0.478(4)(13)$    \\ \cline{3-5} 
    \multicolumn{1}{c|}{}                                      & \multirow{2}{*}{\rotatebox{90}{$\run{2}$}} & \multicolumn{1}{|c|}{$g$ } & $5.67(10)(58)$     & $6.40(16)(24)$     \\
    \multicolumn{1}{c|}{}                                      &                                           & \multicolumn{1}{|c|}{$am$} & $0.5183(15)(50)$   & $0.4763(37)(23)$  \\ \cline{3-5}
    \multicolumn{1}{c|}{}                                      & \multirow{2}{*}{\rotatebox{90}{$\run{3}$}} & \multicolumn{1}{|c|}{$g$ } & $5.58(9)(48)$      & $6.44(15)(33)$    \\
    \multicolumn{1}{c|}{}                                      &                                           & \multicolumn{1}{|c|}{$am$} & $0.5186(10)(43)$   & $0.4843(40)(96)$  \\ \cline{3-5}
    \multicolumn{1}{c|}{}                                      & \multirow{2}{*}{\rotatebox{90}{$\run{4}$}} & \multicolumn{1}{|c|}{$g$ } & $5.63(9)(55)$      & $6.17(16)(47)$   \\
    \multicolumn{1}{c|}{}                                      &                                           & \multicolumn{1}{|c|}{$am$} & $0.5188(10)(46)$   & $0.4640(40)(88)$  \\ \hline
    \multicolumn{2}{c|}{model}                                 & {$g$ }  & $5.64(9)(55)$        & $6.33(15)(31)$    \\
    \multicolumn{2}{c|}{avg.}                                  & {$am$}  & $0.5189(10)(45)$     & $0.4716(36)(70)$      \\\hline\hline

    \multicolumn{1}{c|}{\multirow{8}{*}{\rotatebox{90}{$\ERE$}}} & \multirow{2}{*}{\rotatebox{90}{$\run{1}$}} & \multicolumn{1}{|c|}{$a^{-3}a_1$} & $-28.0(1.2)(5.9)$  & $-32.8(1.3)(6.9)$ \\
    \multicolumn{1}{c|}{}                                      &                                           & \multicolumn{1}{|c|}{$ar_1$     } & $2.61(11)(60)$     & $1.21(10)(38)$    \\ \cline{3-5} 
    \multicolumn{1}{c|}{}                                      & \multirow{2}{*}{\rotatebox{90}{$\run{2}$}} & \multicolumn{1}{|c|}{$a^{-3}a_1$} & $-28.6(1.2)(6.7)$  & $-32.1(1.4)(5.2)$ \\
    \multicolumn{1}{c|}{}                                      &                                           & \multicolumn{1}{|c|}{$ar_1$     } & $2.62(11)(65)$     & $1.21(9)(29)$     \\ \cline{3-5}
    \multicolumn{1}{c|}{}                                      & \multirow{2}{*}{\rotatebox{90}{$\run{3}$}} & \multicolumn{1}{|c|}{$a^{-3}a_1$} & $-27.0(1.1)(5.4)$  & $-29.2(1.5)(1.4)$ \\
    \multicolumn{1}{c|}{}                                      &                                           & \multicolumn{1}{|c|}{$ar_1$     } & $2.73(11)(57)$     & $1.33(13)(6)$     \\ \cline{3-5}
    \multicolumn{1}{c|}{}                                      & \multirow{2}{*}{\rotatebox{90}{$\run{4}$}} & \multicolumn{1}{|c|}{$a^{-3}a_1$} & $-27.6(1.1)(6.1)$  & $-31.4(1.5)(3.6)$ \\
    \multicolumn{1}{c|}{}                                      &                                           & \multicolumn{1}{|c|}{$ar_1$     } & $2.66(10)(64)$     & $1.26(12)(17)$    \\ \hline
    \multicolumn{2}{c|}{model}                                 & $a^{-3}a_1$  & $-27.9(1.1)(6.0)$     & $-33.3(1.3)(6.4)$    \\
    \multicolumn{2}{c|}{avg.}                                  & $ar_1$       & $2.65(11)(61)$        & $1.17(10)(33)$      \\
    \end{tabular}
    }}%
    \caption{Symmetrized model-averaged phase-shift parameters for the ${K\pi \ (I=1/2)}$ and ${\pi\pi \ (I=1)}$ channels, individually computed for each model from Sec.~\ref{sec:phaseshiftdetermination}, and for each choice of hyperparameters (``$\runs$'') as in Table~\ref{tab:hyperparametervariation}. The model-averaged results for each phase-shift model over the hyperparameter variations are also given.}
    \label{tab:phaseshiftpars}
\end{table}

\begin{table}[!h]
    \centering
    \setlength{\tabcolsep}{0.5em}{\renewcommand{\arraystretch}{1.5}{
    \begin{tabular}{ccc|c|c}
        \cline{4-5}
        & &     & $K\pi$ $(I=1/2)$         & $\pi\pi$ $(I=1)$  \\ \hline\hline
        \multicolumn{1}{c|}{\multirow{8}{*}{\rotatebox{90}{$\BW$}}}  & \multirow{2}{*}{\rotatebox{90}{$\run{1}$}} & \multicolumn{1}{|c|}{$aM$}         & $0.5171(9)(42)$   & $0.467(3)(11)$    \\
        \multicolumn{1}{c|}{}                                      &                                           & \multicolumn{1}{|c|}{$a\Gamma$}    & $0.0293(10)(59)$  & $0.102(5)(13)$    \\ \cline{3-5} 
        \multicolumn{1}{c|}{}                                      & \multirow{2}{*}{\rotatebox{90}{$\run{2}$}} & \multicolumn{1}{|c|}{$aM$}         & $0.5161(10)(46)$  & $0.4653(29)(37)$    \\
        \multicolumn{1}{c|}{}                                      &                                           & \multicolumn{1}{|c|}{$a\Gamma$}    & $0.0293(10)(62)$  & $0.1030(55)(62)$  \\ \cline{3-5}
        \multicolumn{1}{c|}{}                                      & \multirow{2}{*}{\rotatebox{90}{$\run{3}$}} & \multicolumn{1}{|c|}{$aM$}         & $0.5165(9)(41)$   & $0.4726(31)(70)$    \\
        \multicolumn{1}{c|}{}                                      &                                           & \multicolumn{1}{|c|}{$a\Gamma$}    & $0.0286(10)(51)$  & $0.107(6)(13)$  \\ \cline{3-5}
        \multicolumn{1}{c|}{}                                      & \multirow{2}{*}{\rotatebox{90}{$\run{4}$}} & \multicolumn{1}{|c|}{$aM$}         & $0.5165(9)(44)$   & $0.4555(30)(70)$   \\
        \multicolumn{1}{c|}{}                                      &                                           & \multicolumn{1}{|c|}{$a\Gamma$}    & $0.0292(9)(59)$   & $0.094(6)(15)$  \\ \hline

        \multicolumn{2}{c|}{model}                                 & $aM$       & $0.5167(9)(43)$     & $0.4626(28)(65)$    \\
        \multicolumn{2}{c|}{avg.}                                  & $a\Gamma$  & $0.0292(10)(59)$     & $0.099(5)(10)$      \\\hline\hline

        \multicolumn{1}{c|}{\multirow{8}{*}{\rotatebox{90}{$\ERE$}}} & \multirow{2}{*}{\rotatebox{90}{$\run{1}$}} & \multicolumn{1}{|c|}{$aM$}         & $0.5168(14)(48)$  & $0.4560(40)(44)$ \\
        \multicolumn{1}{c|}{}                                      &                                           & \multicolumn{1}{|c|}{$a\Gamma$}    & $0.0309(14)(73)$  & $0.140(9)(40)$    \\ \cline{3-5} 
        \multicolumn{1}{c|}{}                                      & \multirow{2}{*}{\rotatebox{90}{$\run{2}$}} & \multicolumn{1}{|c|}{$aM$}         & $0.5155(14)(48)$  & $0.4563(32)(66)$ \\
        \multicolumn{1}{c|}{}                                      &                                           & \multicolumn{1}{|c|}{$a\Gamma$}    & $0.0311(13)(80)$  & $0.136(9)(30)$     \\ \cline{3-5}
        \multicolumn{1}{c|}{}                                      & \multirow{2}{*}{\rotatebox{90}{$\run{3}$}} & \multicolumn{1}{|c|}{$aM$}         & $0.5157(14)(47)$  & $0.4573(54)(26)$ \\
        \multicolumn{1}{c|}{}                                      &                                           & \multicolumn{1}{|c|}{$a\Gamma$}    & $0.0293(12)(63)$  & $0.121(13)(3)$     \\ \cline{3-5}
        \multicolumn{1}{c|}{}                                      & \multirow{2}{*}{\rotatebox{90}{$\run{4}$}} & \multicolumn{1}{|c|}{$aM$}         & $0.5159(13)(49)$  & $0.4535(47)(15)$ \\
        \multicolumn{1}{c|}{}                                      &                                           & \multicolumn{1}{|c|}{$a\Gamma$}    & $0.0307(12)(71)$  & $0.126(11)(14)$    \\ \hline
        \multicolumn{2}{c|}{model}                                 & $aM$       & $0.5158(14)(49)$     & $0.4573(39)(57)$    \\
        \multicolumn{2}{c|}{avg.}                                  & $a\Gamma$  & $0.0305(13)(71)$     & $0.143(9)(37)$      \\
    \end{tabular}
    }}%
    \caption{Symmetrized model-averaged pole position results for the ${K\pi \ (I=1/2)}$ and ${\pi\pi \ (I=1)}$ channels, individually computed for each phase-shift model from Sec.~\ref{sec:phaseshiftdetermination}, and for each choice of hyperparameters (``$\runs$'') as in Table~\ref{tab:hyperparametervariation}. The corresponding overall model-averaged pole-position results are also given in Eqs.~\eqref{overallkstarlattice},\eqref{overallrholattice}.}
    \label{tab:individualpolepos}
\end{table}

\clearpage

\bibliographystyle{apsrev4-1}
\bibliography{scattering}

\end{document}